\documentclass[fleqn,usenatbib]{mnras}

\usepackage[T1]{fontenc}

\DeclareRobustCommand{\VAN}[3]{#2}
\let\VANthebibliography\thebibliography
\def\thebibliography{\DeclareRobustCommand{\VAN}[3]{##3}\VANthebibliography}

\usepackage{graphicx}	
\usepackage{amsmath}	
\usepackage{amssymb}	
\usepackage{multirow}
\usepackage{booktabs}

\title[Modern and Prebiotic Earth Scenarios for TRAPPIST-1e]{Differentiating Modern and Prebiotic Earth Scenarios for TRAPPIST-1e: High-resolution Transmission Spectra and Predictions for \textit{JWST}}

\author[Lin et al.]{
Zifan Lin$^{1,2}$\thanks{E-mail: zl433@cornell.edu},
Ryan J. MacDonald$^{1}$,
Lisa Kaltenegger$^{1}$,
and David J. Wilson$^{3}$
\\
$^{1}$Carl Sagan Institute and Department of Astronomy, Cornell University, Ithaca, NY 14853, USA \\
$^{2}$Department of Earth, Atmospheric and Planetary Sciences, Massachusetts Institute of Technology, Cambridge, MA 02139, USA \\
$^{3}$McDonald Observatory, University of Texas at Austin, Austin, TX 78712, USA
}

\date{Accepted XXX. Received YYY; in original form ZZZ}

\pubyear{2021}

\begin{document}
\label{firstpage}
\pagerange{\pageref{firstpage}--\pageref{lastpage}}
\maketitle

\begin{abstract} 
The TRAPPIST-1 system is a priority target for terrestrial exoplanet characterization. TRAPPIST-1e, residing in the habitable zone, will be observed during the \textit{JWST} GTO Program. Here, we assess the prospects of differentiating between prebiotic and modern Earth scenarios for TRAPPIST-1e via transmission spectroscopy. Using updated TRAPPIST-1 stellar models from the Mega-MUSCLES survey, we compute self-consistent model atmospheres for a 1\,bar prebiotic Earth scenario and two modern Earth scenarios (1 and 0.5\,bar eroded atmosphere). Our modern and prebiotic high-resolution transmission spectra ($0.4 - 20\,\micron$ at $R \sim$100,000) are made available online. We conduct a Bayesian atmospheric retrieval analysis to ascertain the molecular detectability, abundance measurements, and temperature constraints achievable for both scenarios with \textit{JWST}. We demonstrate that \textit{JWST} can differentiate between our prebiotic and modern Earth scenarios within 20 NIRSpec Prism transits via CH$_4$ abundance measurements. However, \textit{JWST} will struggle to detect O$_3$ for our modern Earth scenario to $> 2\,\sigma$ confidence within the nominal mission lifetime ($\sim$ 80 transits over 5 years). The agnostic combination of N$_2$O and/or O$_3$ offers better prospects, with a predicted detection significance of $2.7\,\sigma$ with 100 Prism transits. We show that combining MIRI LRS transits with Prism data provides little improvement to atmospheric constraints compared to observing additional Prism transits. Though biosignatures will be challenging to detect for TRAPPIST-1e with \textit{JWST}, the abundances for several important molecules -- CO$_2$, CH$_4$, and H$_2$O -- can be measured to a precision of $\lesssim$ 0.7 dex (a factor of 5) within a 20 Prism transit \textit{JWST} program. 
\end{abstract}




\section{Introduction}

A primary goal of exoplanetary science is to characterize the atmospheres of temperate rocky planets around other stars. Of the over 4,000 exoplanets across more than 3,000 planetary systems discovered to date, a fraction of these worlds are already known to orbit in the temperate, so called habitable zone (HZ). Such a location may allow for liquid water on the surface of rocky planets (see e.g. \citealt{1993Icar..101..108K}), rendering them intriguing targets for biosignature searches (see the reviews of \citealt{kaltenegger_how_2017}, \citealt{Fujii2018}) with upcoming telescopes. The near-term feasibility of characterizing temperate terrestrial exoplanet atmospheres is largely confined to transiting rocky planets in the HZ of cool M stars, due to their high planet-to-star radius ratio, frequent transits, and their inherently high occurrence rate (see e.g. \citealt{scalo_m_2007}; \citealt{dressing_occurrence_2015}; \citealt{Johns_2018ApJS..239...14J}; \citealt{Berger_2020AJ....159..280B}; \citealt{Bryson2020}). Amongst the M star systems known to host rocky planets in the HZ, the nearby TRAPPIST-1 system (\citealt{gillon_temperate_2016}, \citeyear{Gillon2017}) offers the remarkable opportunity to characterize seven Earth-size transiting planets, four of which orbit in the star's HZ.

TRAPPIST-1e has emerged as a prime candidate for an Earth-like rocky planet. With a radius of 0.918 R$_{\earth}$ and mass of 0.772 M$_{\earth}$ \citep{Gillon2017,Grimm2018}, TRAPPIST-1e has a similar density to the Earth. Due to its relatively large size compared to its star (0.117 R$_{\sun}$, \citealt{Gillon2017}), TRAPPIST-1e displays large transit depths of $\approx 0.5$ per cent. TRAPPIST-1e's close-in orbit ($\approx 0.03$\, au) leads it to receive about 66 per cent of Earth’s irradiation, with transits occurring every 6 days. 

Several groups have discussed the habitability of TRAPPIST-1e from perspectives including atmospheric compositions and escape, surface vegetation, and water inventories, with models ranging from modern Earth-like O$_2$- and CO$_2$-containing atmospheres to Venus-like H$_2$SO$_4$-rich atmospheres (e.g. \citealt{morley_observing_2017}; \citealt{Krissansen-Totton2018}; \citealt{lincowski_evolved_2018}; \citealt{Lustig-Yaeger2019}; \citealt{omalley-james_lessons_2019}; \citealt{Lin2020}). However, due to intense and frequent ultraviolet (UV) flaring activities of M stars, several groups have raised concerns about the surface habitability of planets like TRAPPIST-1e (see e.g. \citealt{scalo_m_2007}; \citealt{shields_habitability_2016}; \citealt{kaltenegger_how_2017}; \citealt{gunther_stellar_2020}). Stellar flares, especially during the early life on a planet, may lead to erosion and water loss that prevent the planet from retaining an atmosphere over geologic timescales (e.g. \citealt{lammer_coronal_2007}; \citealt{see_effects_2014}; \citealt{lingam_reduced_2017}; \citealt{dong_is_2017}). However, new research suggests that a planet can retain water by sequestering it in the mantle \citep{Moore2020}. Despite high amount of top-of-atmosphere UV radiation received by planets orbiting active M stars, surface UV levels are expected to be similar to UV surface levels for a young Earth even for planets with eroded atmospheres orbiting active M stars (\citealt{O'Malley-James_2017MNRAS.469L..26O}, \citeyear{omalley-james_lessons_2019}). Regardless of the true end-state of TRAPPIST-1e's atmosphere, model transmission spectra generally suggest that a wide variety of temperate atmospheres, including Earth-like and non-Earth-like conditions, are on the cusp of detectability with upcoming instruments (see \citealt{Gillon_2020arXiv200204798G} for a recent review).

Existing observations of TRAPPIST-1e have already ruled out hydrogen-dominated atmospheres. Near-infrared \textit{Hubble Space Telescope} (\textit{HST}) transmission spectra of TRAPPIST-1e obtained with the WFC3 instrument (1.1-1.7 $\micron$) display no evidence of the high-amplitude absorption features expected for a cloud-free hydrogen-dominated atmosphere \citep{de_wit_atmospheric_2018}. The required properties for clouds or hazes to produce muted transmission spectra disfavour an aerosol-rich hydrogen-dominated atmosphere \citep{moran_limits_2018}. Combining these spectral inferences with bulk density measurements and atmospheric escape and gas accretion models, the most likely scenario is that TRAPPIST-1e possesses a high mean molecular weight atmosphere \citep{Turbet2020}. However, the current state-of-the-art combination of \textit{\textit{HST}}, \textit{Spitzer}, and ground-based transmission spectra for TRAPPIST-1e does not yet reach the sensitivity to differentiate between classes of terrestrial atmospheres \citep{Ducrot2020}.

The launch of the \textit{James Webb Space Telescope} (\textit{JWST}) and the construction of Extremely Large Telescopes (ELTs) will lift the veil on terrestrial exoplanet atmospheres. The atmospheric detectability of the TRAPPIST-1 planets with \textit{JWST}'s low to medium resolution spectroscopy has seen extensive prior study (e.g. \citealt{Barstow_2016MNRAS.461L..92B}; \citealt{morley_observing_2017}; \citealt{Batalha2018}; \citealt{Krissansen-Totton2018}; \citealt{Lustig-Yaeger2019}; \citealt{wunderlich_detectability_2019}; \citealt{Fauchez2020}). For TRAPPIST-1e, previous studies have indicated that gases such as CO$_2$, H$_2$O, and CH$_4$ can, depending on the atmospheric composition, be detected within $\approx$ 10 transits with \textit{JWST}, while O$_3$ may require at least 100 transits \citep{Gillon_2020arXiv200204798G}. Complementary observations at high spectral resolution with ELTs may permit the detection of spectral features that are difficult to resolve with \textit{JWST}, such as the O$_2$ \textit{A}-band \citep{Snellen2013,Rodler2014,Serindag2019,Hawker2019}. With a sufficient quantity of invested observing time, across many transits and with different instruments, one may anticipate transmission spectra of sufficient quality to infer precise atmospheric properties.

Atmospheric retrieval, or spectral inversion, is a common approach used to extract atmospheric properties from exoplanet spectra \citep[see][for a review]{madhusudhan2018}. Retrieval codes generate $\sim$ millions of model transmission spectra to identify the range of atmospheric properties consistent with a given observed spectrum. Their use of Bayesian sampling algorithms to explore the model parameter space yields two products: (i) posterior distributions for atmospheric properties (e.g. molecular mixing ratios, temperature profiles, cloud properties); and (ii) detection significances for model components from Bayesian model comparisons. Bayesian model comparisons yield robust molecular detection significances that automatically account for information across the full wavelength range and degeneracies between model parameters (e.g. overlapping absorption features) -- offering a distinct advantage over estimating detection confidences from signal-to-noise ratios. For these reasons, retrievals have become the workhorse for inferring atmospheric properties from \textit{HST} exoplanet transmission spectra \citep[e.g.,][]{Wakeford2017,Benneke2019,Lewis2020} and will similarly serve as the bridge to interpret the first \textit{JWST} observations of terrestrial exoplanets.

To date, most retrieval studies of TRAPPIST-1e have considered simplified Earth-like atmospheric compositions that do not account for how the non Sun-like spectral energy distribution (SED) alters the atmospheric chemistry. \citet{Krissansen-Totton2018} assigned proxy molecular abundances inspired by the modern Earth and Archean Earth (assumed constant in altitude) to investigate constraints on retrieved CO$_2$ and CH$_4$ abundances. Similarly, \citet{Tremblay2020} derived predicted detection significances and mixing ratio constraints after assuming a uniform-in-altitude composition modelled on the modern Earth. However, TRAPPIST-1's SED is shifted to longer wavelengths than the Sun, resulting in a higher efficiency of heating the surface of a mostly N$_2$-H$_2$O-CO$_2$ atmosphere due to the decreasing effectiveness of Rayleigh scattering as the star’s spectral peak shifts to longer wavelengths (see e.g. \citealt{1993Icar..101..108K}). The unique UV environment of TRAPPIST-1e also imparts important photochemical changes \citep[e.g.][]{wunderlich_distinguishing_2020, Lin2020}, such that self-consistent simulations of TRAPPIST-1e's atmosphere over its evolutionary history provide a more realistic baseline to assess the prospects for atmospheric retrievals with \textit{JWST}.

\newpage

In this paper, we demonstrate that classes of self-consistent Earth-like atmospheres for TRAPPIST-1e can be differentiated. Specifically, we present atmospheric models and high-resolution transmission spectra for TRAPPIST-1e with a modern biogenic gas flux (`modern Earth') and prior to the emergence of life (`prebiotic Earth'), calculated self-consistently with the recently available semi-empirical model SED of TRAPPIST-1 from the Mega-MUSCLES \textit{HST} treasury program (Wilson et al. 2021, submitted). We provide our simulated high-resolution transmission spectra (0.4--20 $\micron$ at $R \gtrsim 100,000$) \href{http://doi.org/10.5281/zenodo.4770258}{online} to aid future observational and theoretical simulations of TRAPPIST-1e. Finally, we conduct an extensive atmospheric retrieval analysis to provide predicted constraints on TRAPPIST-1e's atmosphere, for different sized \textit{JWST} programs for both these self-consistent scenarios, using the Mega-MUSCLES TRAPPIST-1 SED to enable realistic \textit{JWST} noise simulations.

In what follows, we describe the Mega-MUSCLES TRAPPIST-1 spectrum in Section~\ref{sec:MUSCLES_spectrum}. This stellar spectrum is the stellar input to our TRAPPIST-1e atmosphere models in Section~\ref{sec:atmospheric_models}. We present our high-resolution TRAPPIST-1e transmission spectra in Section~\ref{sec:high_res_spectra}. We explore the prospect of differentiating modern Earth and prebiotic Earth scenarios with \textit{JWST} in Section~\ref{sec:JWST_predictions}. Finally, in Section~\ref{sec:summary_discussion}, we summarize our results and discuss the implications.

\section{A Mega-MUSCLES Stellar Model for TRAPPIST-1} \label{sec:MUSCLES_spectrum}

Mega-MUSCLES (Measurements of the Ultraviolet Spectral Characteristics of Low-Mass Exoplanetary Systems) is an \textit{HST} Treasury program obtaining 5\,\AA--10\,$\micron$ SEDs of a sample of 13 M\,dwarfs, including TRAPPIST-1, covering a wide range of stellar mass, age, and planetary system architectures. It extends the original 11-star MUSCLES program (\citealt{france_muscles_2016}; \citealt{Youngblood_2016ApJ...824..101Y}; \citealt{loyd_muscles_2018}) to stars with lower masses, higher activity, and/or faster rotation rates. 

The MUSCLES spectroscopic observations of TRAPPIST-1's SED combine UV and blue-optical spectroscopy obtained with the Hubble Space Telescope (1130--5700\,\AA), X-ray spectroscopy obtained with XMM-Newton (5--50\,\AA), and models of the stellar photosphere, chromosphere, transition region and corona. The combined SED is shown in Figure~\ref{fig:stellar_spectra}. The low SNR \textit{HST} data at UV wavelengths is replaced by a semi-empirical model, comprised of model fits to the observed emission lines combined with an underlying continuum model. PHOENIX models from the latest BT-Settl/CIFIST grid\footnote{\url{http://svo2.cab.inta-csic.es/theory//newov2/index.php?models=bt-settl-cifist}} (\citealt{Baraffe_2015A&A...577A..42B}; \citealt{Allard_2016sf2a.conf..223A}) cover the optical and infrared wavelengths. As with the MUSCLES SEDs, the Lyman $\alpha$ line was reconstructed from \textit{HST} data via the methods presented in \citet{Youngblood_2016ApJ...824..101Y} and the X-ray data fitted with an APEC model spectrum. The extreme UV region (120--1000\,\AA) is represented with a state-of-the-art Differential Emission Measure model (Duvvuri et al., submitted).   A full description of the TRAPPIST-1 observations and SED production is presented in \citet{Wilson2021}.

We calculated the flux reaching TRAPPIST-1e by scaling the MUSCLES TRAPPIST-1 spectrum observed at Earth by the squared ratio of the distance to the star divided by the distance to the planet. We used a distance of 12.425 pc, derived from the GAIA DR2 parallax \citep{Bailer-Jones_2018AJ....156...58B}, and an orbital distance of $0.02817$ au for TRAPPIST-1e \citep{Gillon2017}.

\section{Atmospheric Models of TRAPPIST-1e} \label{sec:atmospheric_models}

The atmosphere composition of terrestrial planets depend on the irradiation from its host star, surface type, cloud coverage, outgassing rates, subduction rates of individual chemicals, and subsequent photochemistry.

We model three different types of atmosphere for TRAPPIST-1e: (i) a 1 bar modern Earth scenario, i.e. a planet with modern Earth outgassing rates (see e.g. \citealt{Rugheimer_2013AsBio..13..251R}), a 1 bar surface pressure, and a CO$_2$ abundance 100 times present atmospheric levels (PAL) to ensure surface temperatures above freezing (see section~\ref{subsec:modern_prebiotic_scenarios}); (ii) a 0.5 bar modern Earth scenario, the same as scenario (i) but with a lower surface pressure to explore the effect of surface pressure on the composition and transmission spectrum of an eroded atmosphere; and (iii) a prebiotic Earth scenario, with increased CO$_2$ (10 per cent mixing ratio) and only trace quantities of free oxygen, corresponding to Earth’s atmosphere before the Great Oxidation Event (see 
\citealt{lyons_rise_2014}, \citealt{zahnle_emergence_2007}, and \citealt{kaltenegger_high-resolution_2020}). 

The stellar and planetary parameters used in our models are taken from \cite{Gillon2017} and \cite{Grimm2018} unless otherwise noted. We recognize that more recent estimates of these parameters are available \citep{Agol2021PSJ.....2....1A}, but the older values are used for model consistency.

\subsection{Self-consistent Atmospheric Modelling Procedure}

We use Exo-Prime2 (described in detail in \citealt{Madden_2020_surfaces}), a one-dimensional rocky exoplanet atmosphere code that couples a climate and photochemistry model to a line-by-line radiative transfer code. The code was originally developed to retrieve trace gases in Earth's atmosphere \citep{Traub_1976ApOpt..15..364T} and later adapted to simulate emergent and transmission exoplanet spectra (e.g. \citealt{kaltenegger_spectral_2007}, \citeyear{kaltenegger_transits_2009}). Visible and near-IR shortwave fluxes are calculated with a two-stream approximation including atmospheric gas scattering \citep{Toon_1989JGR....9416287T}, while longwave fluxes in the IR region are calculated with a rapid radiative transfer model. Exo-Prime2 also uses a wavelength dependent surface albedo for clouds and planetary surfaces. We model the 1D global atmospheric profiles using a plane-parallel atmosphere, setting the stellar zenith angle to 60$^\circ$ to represent the average incoming stellar flux on the dayside of the planet (see also \citealt{Schindler_2000Icar..145..262S}). The photochemistry code (originally developed by \citealt{Kasting_1986Sci...234.1383K}) contains 220 reactions to solve for 55 chemical species and uses a reverse-Euler solver. 

\subsection{Modern and Prebiotic Earth Scenarios: Atmospheric Compositions for TRAPPIST-1e} \label{subsec:modern_prebiotic_scenarios}

\begin{figure}
	\includegraphics[width=\columnwidth]{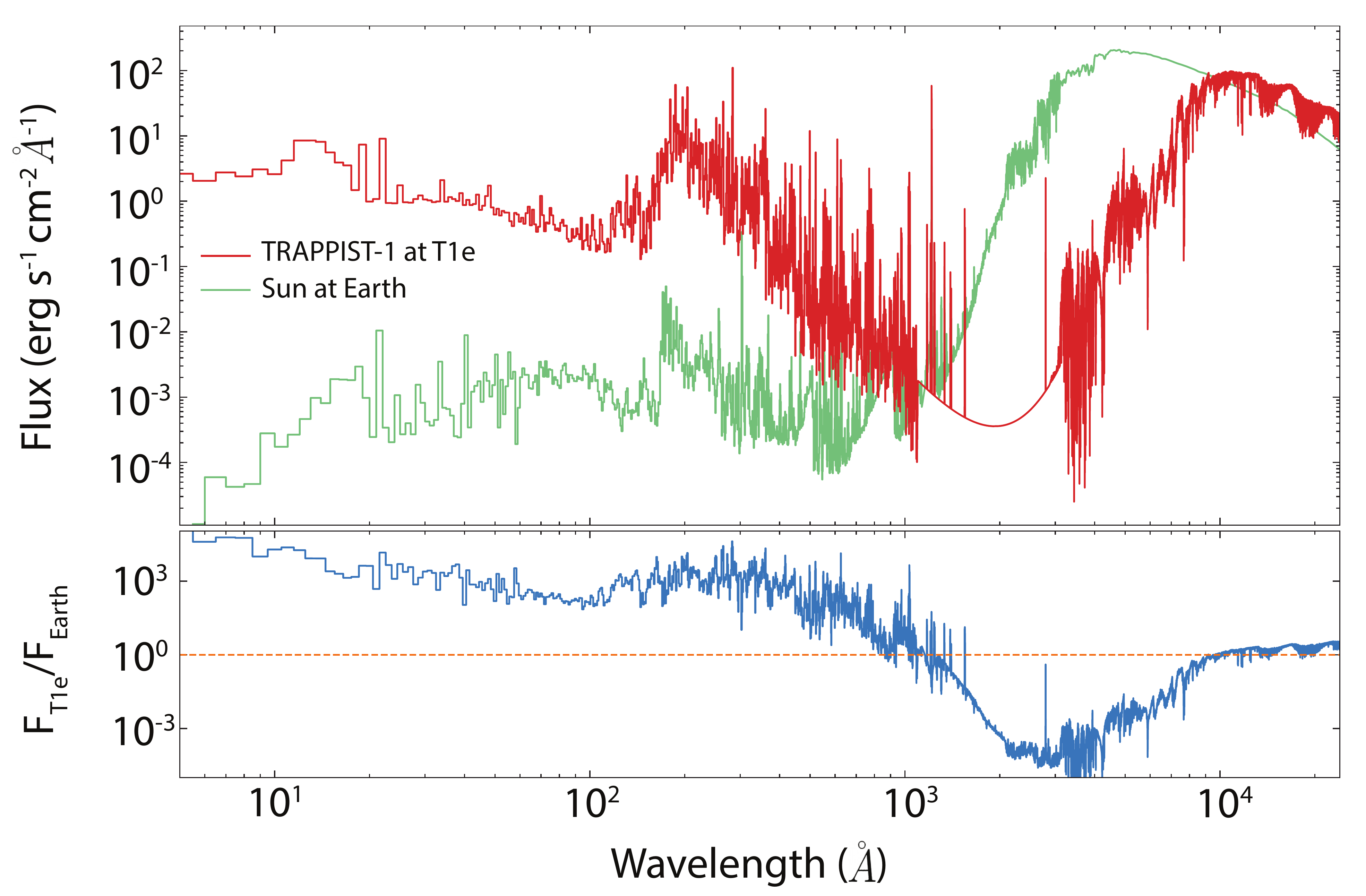}
    \caption{The Mega-MUSCLES SED of TRAPPIST-1 compared with the SED of the modern Sun (upper panel). The TRAPPIST-1 SED was scaled to the distance of TRAPPIST-1e using the Gaia DR2 parallax \citep{Gaia_Collaboration_2018} and the semi-major axis of TRAPPIST-1e \citep{Gillon2017}. The solar spectrum was scaled to 1 AU. The lower panel shows the ratio of irradiation received by TRAPPSIT-1e compared to Earth's irradiation. TRAPPIST-1e receives $\sim10^3$ more X-ray/Extreme UV irradiation compared to modern Earth.}
    \label{fig:stellar_spectra}
\end{figure}

\begin{table*}
\caption{Surface temperature and mixing ratios of major atmospheric components in our model atmospheres. The modern TRAPPIST-1e models represent atmospheres with outgassing rates equal to those of the modern Earth, but with 100$\times$ the PAL CO$_{2}$ mixing ratio. The prebiotic TRAPPIST-1e model represents an atmosphere similar to Earth’s prebiotic stage before the rise of oxygen.}
\label{tab:surface_parameters}
\begin{tabular}{lllllllll}
\hline
\multirow{2}{*}{Model}              & \multirow{2}{*}{P$_{\textrm{surf}}$ (bar)} & \multirow{2}{*}{T$_{\textrm{surf}}$ (K)} & \multicolumn{6}{c}{Surface mixing ratio}                     \\
                                    &                              &                            & CO$_2$      & H$_2$O     & O$_2$      & O$_3$       & CH$_4$     & N$_2$O      \\ \hline
\multirow{2}{*}{Modern} & 1.0                          & 289                        & $3.60 \times 10^{-2}$ & $1.36 \times 10^{-2}$ & 0.21    & $3.90 \times 10^{-9}$  & $2.38 \times 10^{-5}$ & $3.74 \times 10^{-7}$  \\
    & 0.5                          & 274                        & $3.60 \times 10^{-2}$ & $1.00 \times 10^{-2}$ & 0.21    & $1.05 \times 10^{-8}$  & $2.53 \times 10^{-5}$ & $3.84 \times 10^{-7}$  \\ \hline
Prebiotic               & 1.0                          & 294                        & 0.1      & $1.84 \times 10^{-2}$ & $1.00 \times 10^{-13}$ & $2.75 \times 10^{-17}$ & $2.20 \times 10^{-6}$ & $5.68 \times 10^{-11}$ \\ \hline
\end{tabular}
\end{table*}

\begin{figure*}
	\includegraphics[width=\textwidth,  trim={0.0cm 0.4cm 0.0cm 0.0cm}]{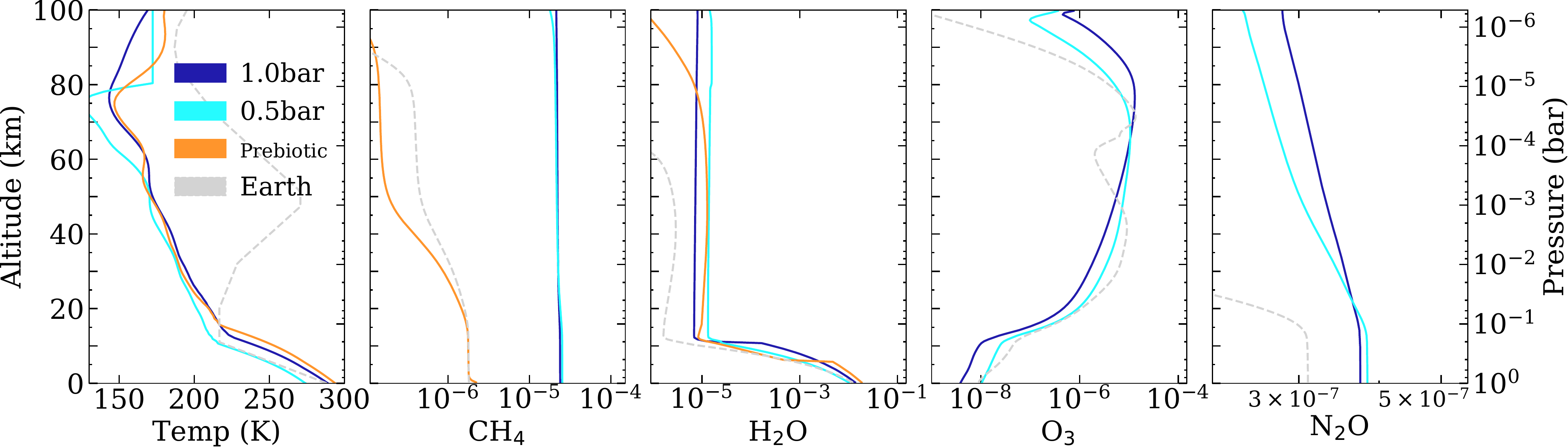}
    \caption{Temperature profiles and mixing ratios of major chemicals in our TRAPPIST-1e atmosphere models. We compare two surface pressures for a modern Earth-like model -- 1 bar (navy) and 0.5 bar (cyan) -- and a prebiotic Earth model (orange). The temperature structure and atmospheric composition for the present day Earth is overplotted comparison (grey dashed profiles). The O$_3$ and N$_2$O mixing ratios are negligible for the prebiotic scenario, so are omitted for clarity. The temperature profile for the 0.5 bar modern model was only computed to about 80 km altitude, so is here extended isothermally to 100 km.}
    \label{fig:atm_profiles}
\end{figure*}

Our initial simulations of TRAPPIST-1e for modern Earth outgassing rates found the planet was not warm enough to produce an average surface temperature above the water freezing point. However, assuming that a geological cycle like the carbonate-silicate cycle on Earth could be maintained on a terrestrial TRAPPIST-1e, the CO$_2$ concentration in the atmosphere should increase and consequently warm the surface of the planet. We therefore increased the mixing ratio of CO$_2$, arbitrarily, to 100 times the PAL (to $3.56 \times 10^{-2}$) to maintain an average surface temperature above freezing. We note that 10$\times$ PAL CO$_2$ leads to a surface temperature just slightly below freezing for a 1 bar surface pressure, while 100$\times$ PAL CO$_2$ leads to a surface pressure above freezing for both 1 bar and 0.5 bar surface pressures. 

We summarize key properties of our TRAPPIST-1e models in Table~\ref{tab:surface_parameters}. We focus on the surface pressure and temperature, and the surface mixing ratios for the most spectroscopically significant chemical species. Figure~\ref{fig:atm_profiles} shows the full temperature profiles and mixing ratio profiles for CH$_4$, H$_2$O, O$_3$, and N$_2$O across our three models. Our atmospheric profiles for TRAPPIST-1e differ significantly from those of Earth (Figure~\ref{fig:atm_profiles}), due to the different SED of its red and cool host star (Figure~\ref{fig:stellar_spectra}). We note that our prebiotic scenario is slightly warmer than the 1 bar modern scenario, despite their identical surface pressures, due to an enhanced greenhouse effect from the higher CO$_2$ mixing ratio in the prebiotic model (10 per cent).

Our three scenarios for TRAPPIST-1e lead to correspondingly different atmospheric compositions. The surface H$_2$O mixing ratio decreases as surface temperature decreases due to lower rate of evaporation. The O$_3$ mixing ratio either decreases or increases with surface pressure, dependent on the altitude, while the O$_3$ mixing ratio of the prebiotic model is negligible. In the upper atmosphere, the CH$_4$ mixing ratio decreases as surface pressure decreases, because the rate of CH$_4$ destruction due to UV photolysis increases in the upper atmosphere. N$_2$O has a larger mixing ratio in our modern scenarios than the present day Earth, and is considered a potential biosignature because its abiotic sources are small in an oxygen-rich atmosphere (see e.g. \citealt{kaltenegger_how_2017}; \citealt{Grenfell_2017PhR...713....1G}; \citealt{Schwieterman_2018AsBio..18..663S}). Both our modern models assume a biological origin for N$_2$O (see \citealt{rugheimer_spectra_2018}), while trace quantities of N$_2$O are photochemical byproducts in our prebiotic model.

\begin{figure*}
	\includegraphics[width=\textwidth,  trim={0.0cm 0.4cm 0.0cm 0.0cm}]{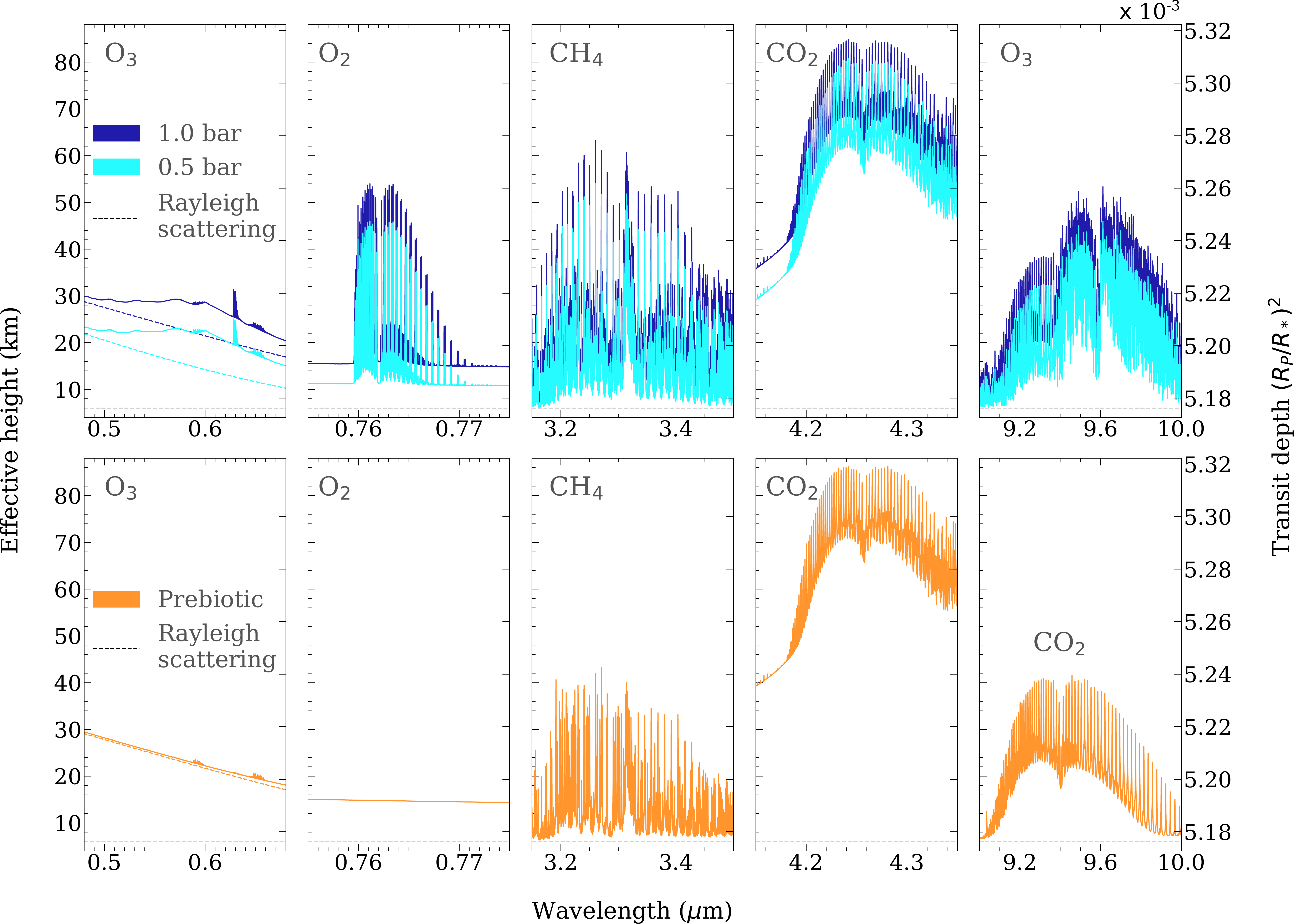}
    \caption{High-resolution spectral features for CO$_2$ and the constituents of the biosignature pairs O$_2$+CH$_4$ and O$_3$+CH$_4$. The model spectra were calculated at a wavenumber resolution of 0.01 cm$^{-1}$, shown here at a uniform resolution of $R$ = 100,000. Top row: our two modern Earth scenarios -- 1 bar surface pressure (dark blue) and 0.5 bar surface pressure (light blue). Bottom row: our prebiotic scenario (orange). Rayleigh scattering slopes are shown in the leftmost panels (dashed lines) to indicate the difference in line shape between the Rayleigh slope and the O$_3$ Chappuis band. The spectra are plotted in two units: (i) effective height (left y-axis), which shows the apparent size of the planet at a certain wavelength due to additional opacity from atmospheric absorption; and (ii) transit depth (right y-axis), which shows $(R_p / R_*)^2$. The main differences between the prebiotic and modern scenarios are the absence of oxygenic molecular features and the decreased CH$_4$ feature strength for the prebiotic scenario. Note that CO$_2$ has a feature around $9.4\,\micron$, which is partially present in the O$_3$ panel for our modern scenarios.}
    \label{fig:highres_features}
\end{figure*}

\begin{figure*}
	\includegraphics[width=\textwidth, trim={0.0cm 0.0cm 0.0cm 0.0cm}]{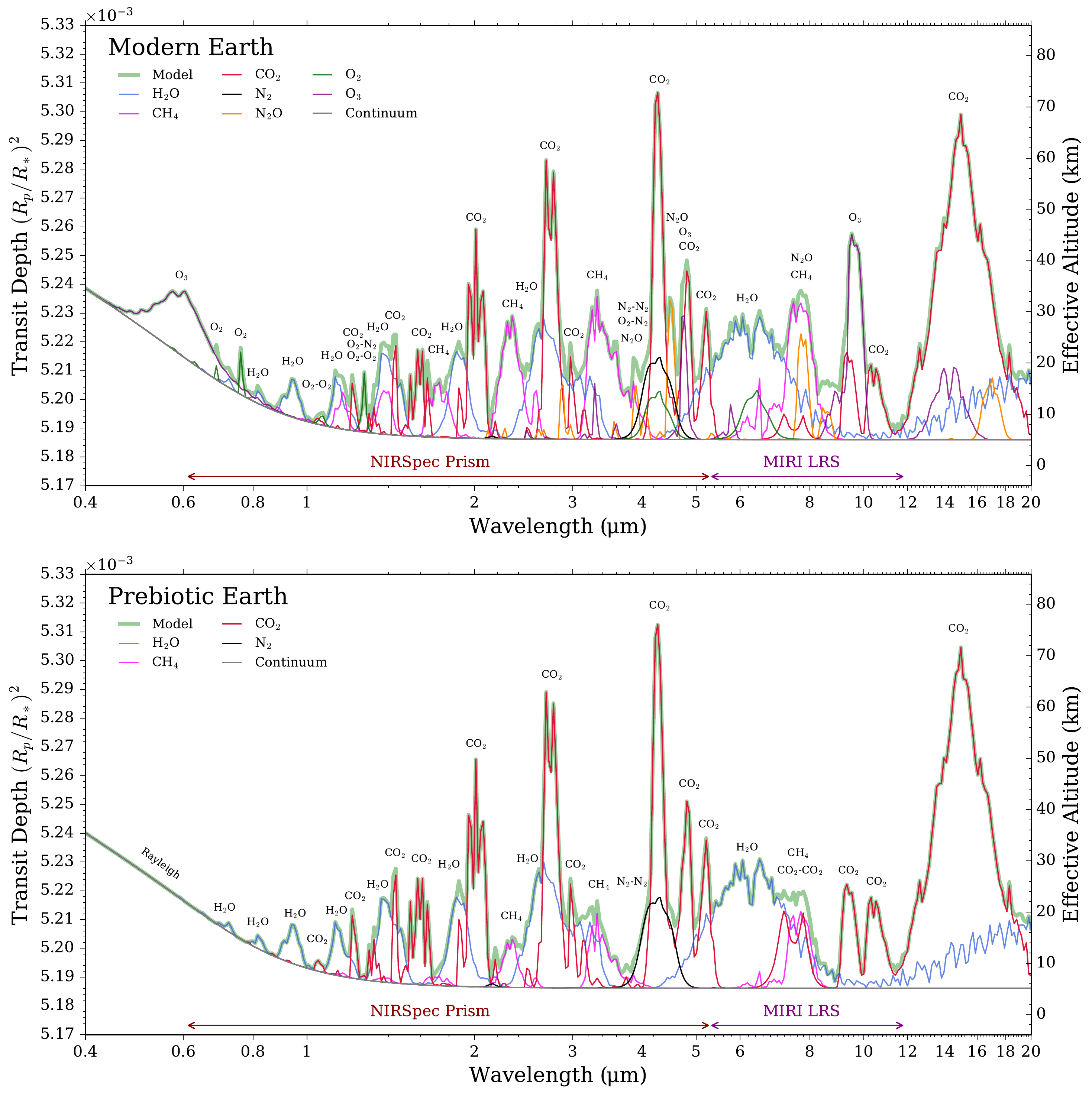}
    \caption{Molecular contributions to model transmission spectra of TRAPPIST-1e. Top: 1\,bar modern Earth scenario. Bottom: prebiotic Earth scenario. The model transmission spectra (green shading) are decomposed into the opacity contributions of each molecule (colored curves). Each molecular contribution is relative to the spectral continuum due to Rayleigh scattering and refraction (gray curve). Prominent absorption features are labeled. Collision-induced absorption (CIA) pairs including O$_2$ (O$_2$-O$_2$ and O$_2$-N$_2$) are depicted alongside O$_2$ line absorption. N$_2$-N$_2$ CIA contributes to both models around $4.3\,\micron$, while CO$_2$-CO$_2$ CIA contributes around $7.5\,\micron$. The equivalent altitude above the surface -- for which an opaque atmosphere would produce the same transit depth at a given wavelength -- is shown on the right. All contributions are plotted at $R=100$ for clarity. The clearest distinguishing feature between the modern and prebiotic Earth scenarios is the lack of O$_3$, O$_2$, and N$_2$O opacity.}
    \label{fig:molecular_signatures}
\end{figure*}

\section{High-resolution Transmission Spectra for TRAPPIST-1e} \label{sec:high_res_spectra}

We use the new UV spectra for TRAPPIST-1 to model high-resolution transmission spectra for the three scenarios described above. In what follows, we describe our radiative transfer modelling in section~\ref{subsec:radiative_transfer} and summarize the key molecular absorption signatures in section~\ref{subsec:molecular_signatures}.

\subsection{Radiative Transfer} \label{subsec:radiative_transfer}

We use the Exo-Prime2 code to generate the high-resolution transmission spectra for our TRAPPIST-1e models from 0.4 to 20 $\micron$ at a resolution of 0.01 cm$^{-1}$ (following \citealt{kaltenegger_transits_2009}). We divide the model atmosphere into 52 layers. For each atmospheric layer line shapes and widths are calculated individually with Doppler- and pressure-broadening with several points per line width. We include the most spectroscopically relevant molecules in our calculations: C$_2$H$_6$, CH$_4$, CO, CO$_2$, H$_2$CO, H$_2$O, H$_2$O$_2$, H$_2$S, HNO$_3$, HO$_2$, N$_2$O, N$_2$O$_5$, NO$_2$, O$_2$, O$_3$. OCS, OH, and SO$_2$, using the HITRAN 2016 line lists \citep{Gordon2017} as well as Rayleigh scattering.

While atmospheric refraction can limit the access to the lower atmosphere of exoplanets through transmission spectroscopy, depending on the spectral type of the host star, the terrestrial planets in the HZ of TRAPPIST-1e are not restricted by refraction (see e.g. \citealt{betremieux_impact_2014}). Therefore, we model the full column of atmosphere from the surface to a top of atmosphere with a minimum pressure of $1.0 \times 10^{-7}$ bar.

Clouds do not significantly affect the strengths of the spectral features in Earth’s transmission because most clouds on Earth are located at altitudes below 12 km, which means they are below the refractive limit of a rocky planet around a Sun-like star. However, for planets that can be probed deeper than 12 km, clouds will start to obscure parts of the spectra. Clouds in transmission spectra will obscure spectral features below the cloud layer, if they occur close or on the terminator region that is probed during primary transit (e.g. \citealt{Seager_2005AsBio...5..372S}; \citealt{kaltenegger_transits_2009}; \citealt{Robinson_2011AsBio..11..393R}; \citealt{betremieux_impact_2014}). Currently, the horizontal and vertical distribution of clouds on Earth-like planets orbiting different host stars is highly uncertain. Thus we explore the effect of clouds by adding one cloud layer in our atmosphere models at 6 km -- the middle layer of Earth clouds, which are located at 1, 6, and 12 km. We consider the pessimistic case for a 100\% opaque cloud layer, uniform around the terminator, in all our models.

\subsection{Molecular Signatures in Transmission Spectra of TRAPPIST-1e} \label{subsec:molecular_signatures}

Here we provide a brief inventory of the most prominent spectral features shaping our model TRAPPIST-1e spectra. Figure~\ref{fig:highres_features} compares absorption lines for several molecules at high-resolution for our modern Earth and prebiotic Earth scenarios. Figure~\ref{fig:molecular_signatures} provides a complementary spectral decomposition into molecular bands at low-resolution.

\subsubsection{Prominent Molecular Absorption Features}

At high-resolution, both our modern Earth scenarios display many strong absorption lines for several molecules. Our modern scenarios show strong spectral features for the O$_2$+CH$_4$ and O$_3$+CH$_4$ biosignature pairs (Figure~\ref{fig:highres_features}, top row) at high-resolution ($R = 100,000$). In comparison, our prebiotic scenario has only trace quantities of oxygenic species (e.g., O$_2$, O$_3$, and N$_2$O), resulting in an optical dominated by Rayleigh scattering and devoid of strong spectral lines (Figure \ref{fig:highres_features}, bottom row). While both scenarios display strong CO$_2$ bands in the infrared, the CH$_4$ features are generally stronger for the modern scenarios due to the enhanced CH$_4$ abundance from biological production. We note that our 1 bar and 0.5 bar spectra are offset because we place the radius of the planetary surface at two different pressures in these forward models. The strength of the absorption features, however, are very similar for both our modern Earth scenarios. Such an offset is typically fit in atmospheric retrieval analyses by either a free planetary radius or reference pressure parameter, so this would not affect the retrieved atmospheric properties. Due to the similar feature strengths for both modern Earth models, in section~\ref{sec:JWST_predictions} we only compare the 1 bar modern model with the prebiotic model.

At low-resolution, the forest of spectral lines merge into molecular bands spanning the optical and infrared. Figure~\ref{fig:molecular_signatures} shows the spectral contributions for the most prominent molecules in our models, spanning our full wavelength range at low-resolution ($R = 100$). Broadly speaking, one would expect the strong CO$_2$ bands to be the first atmospheric feature that would be observed with low-resolution infrared spectra. The most apparent differences between the modern and prebiotic scenarios are: (i) the enhanced CH$_4$ opacity in the near-infrared for the modern scenario, and (ii) the complete absence of O$_3$, O$_2$, and N$_2$O bands in the prebiotic scenario. In what follows, we summarize the locations of the main absorption features for each molecule.

\subsubsection{A Spectral Inventory for TRAPPIST-1e}

The strongest CO$_2$ spectral absorption features are located around 1.4, 1.6, 1.9, 2.1, 2.8, 3.0, 4.2, 4.8, 5.2, 10.3, and 15 $\micron$. Though prominent in all three of our models, the CO$_2$ features are somewhat stronger in our prebiotic scenario than our modern scenario (see Figure~\ref{fig:molecular_signatures}). This arises from the higher CO$_2$ concentration in our prebiotic scenario (10 per cent vs. $3.60 \times 10^{-2}$ for the modern scenarios).

CH$_4$ shows prominent absorption features around 1.7, 2.3, 3.4, and 7.5 $\micron$. The CH$_4$ features are stronger in our modern scenarios than the prebiotic scenario due to the higher modern Earth CH$_4$ mixing ratio. The strength of the CH$_4$ absorption features are similar in both modern scenarios, but the two spectra are offset and the 0.5 bar scenario has a lower effective height, due to the same surface radius assumed for both models as previously mentioned. We note that the CH$_4$ feature at 3.4 $\micron$, if observed in combination with the O$_2$ feature at 0.76 $\micron$, constitutes a `classical' biosignature (see section~\ref{subsubsec:biosignatures}) and the strongest biosignature pair in the visible to near-IR range (0.4--5.0 $\micron$).

The strongest H$_2$O features occur around 0.95, 1.15, 1.32, 1.88, 2.6, 3.75, 6.4, and 17--20 $\micron$. The strength of H$_2$O spectral absorption features are comparable between the 1\,bar modern scenario and the prebiotic scenario. In the visible wavelength range, H$_2$O and CH$_4$ features overlap at 1.15 and 1.32 $\micron$ (see also \citealt{kaltenegger_high-resolution_2020}), but the H$_2$O features are stronger in our models. Similarly, the broad H$_2$O feature at 6.4\,$\micron$ overlaps with two weak CO$_2$ features from 7 to 8 $\micron$. Resolving such overlapping features requires that at least part of their wings are resolved. However, even at the low resolution shown in Figure \ref{fig:molecular_signatures} ($R = 100$) the wings have sufficiently different shapes that they should be distinguishable. Nevertheless, such overlapping features can be more easily disentangled at high-resolution ($R > 100,000$).

For both modern models, O$_2$ shows a strong spectral absorption feature at 0.76 $\micron$ (the A-band) and several features due to collision-induced absorption in the infrared. These O$_2$ features span a relatively limited wavelength range that, when combined with their `sharp' nature (see Figure~\ref{fig:molecular_signatures}), makes their signatures at low-resolution more challenging to identify than at high-resolution.

\newpage

O$_3$ shows a broad continuum feature in the optical from 0.45 to 0.74 $\micron$ (the Chappuis band), alongside a strong infrared feature at 9.6 $\micron$ and a weaker feature at 4.7 $\micron$. These features are absent from the anoxic prebiotic scenario. We note that CO$_2$ has a feature around 9.4 $\micron$ that is partially obscured by the stronger 9.6 $\micron$ O$_3$ feature in both modern models, but these features are sufficiently distinct at both high-resolution (Figure~\ref{fig:highres_features}) and low-resolution (Figure~\ref{fig:molecular_signatures}) to distinguish CO$_2$ from O$_3$.

\newpage

The strongest N$_2$O features are located around 4.5, 7.9 and 17 $\micron$. The detection of N$_2$O features can be challenging, because its 4.5 $\micron$ feature overlaps with CO$_2$ while its 7.9 $\micron$ feature overlaps with CH$_4$, CO$_2$, and H$_2$O. A few weaker N$_2$O features are located around 2.9, 3.6, 3.9, and 4.1 $\micron$, but all are obscured by stronger features in the same wavelength range.

Finally, we note that N$_2$ shows a broad feature centered around 4.3 $\micron$ arising from N$_2$-N$_2$ collision-induced absorption. This feature offers one of the only avenues to constrain N$_2$ abundances from exoplanet transmission spectra \citep{Schwieterman2015}. However, the feature is mostly obscured by the strong CO$_2$ band also located at 4.3 $\micron$.

\subsubsection{Biosignatures in The Modern TRAPPIST-1e Transit Spectra} \label{subsubsec:biosignatures}

\cite{Sagan_1993Natur.365..715S} analyzed the spectrum of Earth -- observed by the Galileo probe -- searching for signatures of life and concluded that the large amount of O$_2$ and the simultaneous presence of CH$_4$ traces are strongly suggestive of biology. O$_2$ or O$_3$ spectral features, in combination with CH$_4$ features, are considered strong biosignatures because CH$_4$ reacts with O$_2$ quickly to produce CO$_2$ and H$_2$O, so both species must be continuously supplied to remain detectable (\citealt{Lovelock_1965Natur.207..568L}; \citealt{Lederberg_1965Natur.207....9L}; \citealt{Sagan_1993Natur.365..715S}). While none of these species can be considered as a biosignature individually, O$_2$+CH$_4$ and O$_3$+CH$_4$ pairs are compelling indicators of a biological origin (e.g. reviews by \citealt{Kasting_2014PNAS..11112641K};  \citealt{kaltenegger_how_2017}; \citealt{Fujii2018}; \citealt{Schwieterman_2018AsBio..18..663S}). Alternatively, N$_2$O has been proposed as a possible biosignature for Earth-like atmospheric conditions (e.g. \citealt{Lederberg_1965Natur.207....9L, Segura_2005AsBio...5..706S, Rugheimer_2013AsBio..13..251R}; \citealt{Grenfell_2014P&SS...98...66G}), but abiotic mechanisms like solar flares can also produce N$_2$O efficiently (e.g. \citealt{airapetian_how_2017}). However, the combination of O$_2$+N$_2$O or O$_3$+N$_2$O can be considered a biosignature (e.g. \citealt{Segura_2005AsBio...5..706S}; \citealt{Rugheimer_2013AsBio..13..251R}). Both the O$_2$+CH$_4$ and O$_2$+N$_2$O biosignature pairs are present in our modern scenario but absent in the prebiotic scenario, which is a major difference between the two scenarios. However, the presence of biosignature pairs in the theoretical high-resolution spectra does not guarantee their detectability, or the differentiability between our modern and prebiotic scenarios, with \textit{JWST}. Therefore, in the next section we simulate \textit{JWST} observations and perform a retrieval analysis to quantitatively address the question of biosignature inferences with \textit{JWST}.

\section{Predictions for \textit{JWST} Transmission Spectroscopy of TRAPPIST-1e} \label{sec:JWST_predictions}

We now turn to quantitative expectations for \textit{JWST} observations of TRAPPIST-1e. Our goal is to establish how many transits of TRAPPIST-1e are required to realistically differentiate modern Earth and prebiotic Earth scenarios (only the 1 bar modern model is considered in this section to ensure fair comparison). We accomplish this via an extensive atmospheric retrieval analysis. In what follows, we describe the simulation of \textit{JWST} transmission spectra in section~\ref{subsec:simulated_data}, our retrieval strategy in section~\ref{subsec:retrieval_methods}, and present our results in section~\ref{subsec:retrieval_results}. 

\subsection{Simulated \textit{JWST} Observations} \label{subsec:simulated_data}

We simulate \textit{JWST} observations of TRAPPIST-1e using PandExo \citep{Batalha2017}. To maximally explore the wavelength coverage of our model transmission spectra, we consider both NIRSpec Prism (0.6--5.3 $\micron$) and MIRI LRS (5--14 $\micron$) observations. For Prism observations, we simulate the partial saturation strategy of \citet{Batalha2018} by adopting 6 groups per integration with the SUB512 subarray, thereby raising the observing efficiency to 71.4\% \citep{Lustig-Yaeger2019}. For MIRI LRS, PandExo's optimiser determined 175 groups per integration and we consider only data below 12\,$\micron$ (due to diminishing precision at longer wavelengths). For both the NIRSpec Prism and MIRI LRS, we limit saturation to 80 per cent full well and keep all simulated data at their native resolution without binning. The input stellar spectrum was the new Mega-MUSCLES TRAPPIST-1 spectrum (section~\ref{sec:MUSCLES_spectrum}), normalised to J = 11.354 \citep{Cutri2003}. The total observing time per transit is taken as three times the transit duration (0.954\,hr; \citealt{Gillon2017}), for 2.86\,hr per transit. With these considerations, we find a single transit mean precision of 510\,ppm for the Prism and 1750\,ppm for MIRI.

We generated simulated \textit{JWST} observations to explore a range of prospective observing campaigns. The \textit{JWST} Guaranteed Time Observations (GTO) for TRAPPIST-1e constitute a 4 transit reconnaissance program with the NIRSpec Prism\footnote{GTO program \href{http://www.stsci.edu/JWST/observing-programs/program-information?id=1331}{\#1331} (PI: Nikole Lewis).}. Our objective is to establish how many additional transits are required to differentiate between modern and prebiotic Earth scenarios. We simulated Prism and MIRI observations, for campaigns ranging from 10 transits through 200 transits, in the following combinations: 10 Prism, 10 Prism + 10 MIRI, 20 Prism, 100 Prism, 100 Prism + 100 MIRI, and 200 Prism. Note that we estimate 81 transits of Trappist-1e in JWSTs nominal 5-year mission \citep{Lustig-Yaeger2019} and assume an extended mission lifetime for the high end of our simulations. For reference, we estimate 6 transits can fit in a small \textit{JWST} campaign ($<$ 25\,hr), 18 transits in a medium campaign ($<$ 75\,hr), and a large campaign is required for additional transits - these assume a 40\% overhead charge, for 4\,hrs of charged time per transit. The combined Prism + MIRI campaigns allow us to investigate O$_3$ abundance constraints via its 9.6\,$\micron$ feature. 

\begin{figure*}
	\includegraphics[width=\textwidth, trim={0.4cm 0.2cm 0.2cm -0.5cm}]{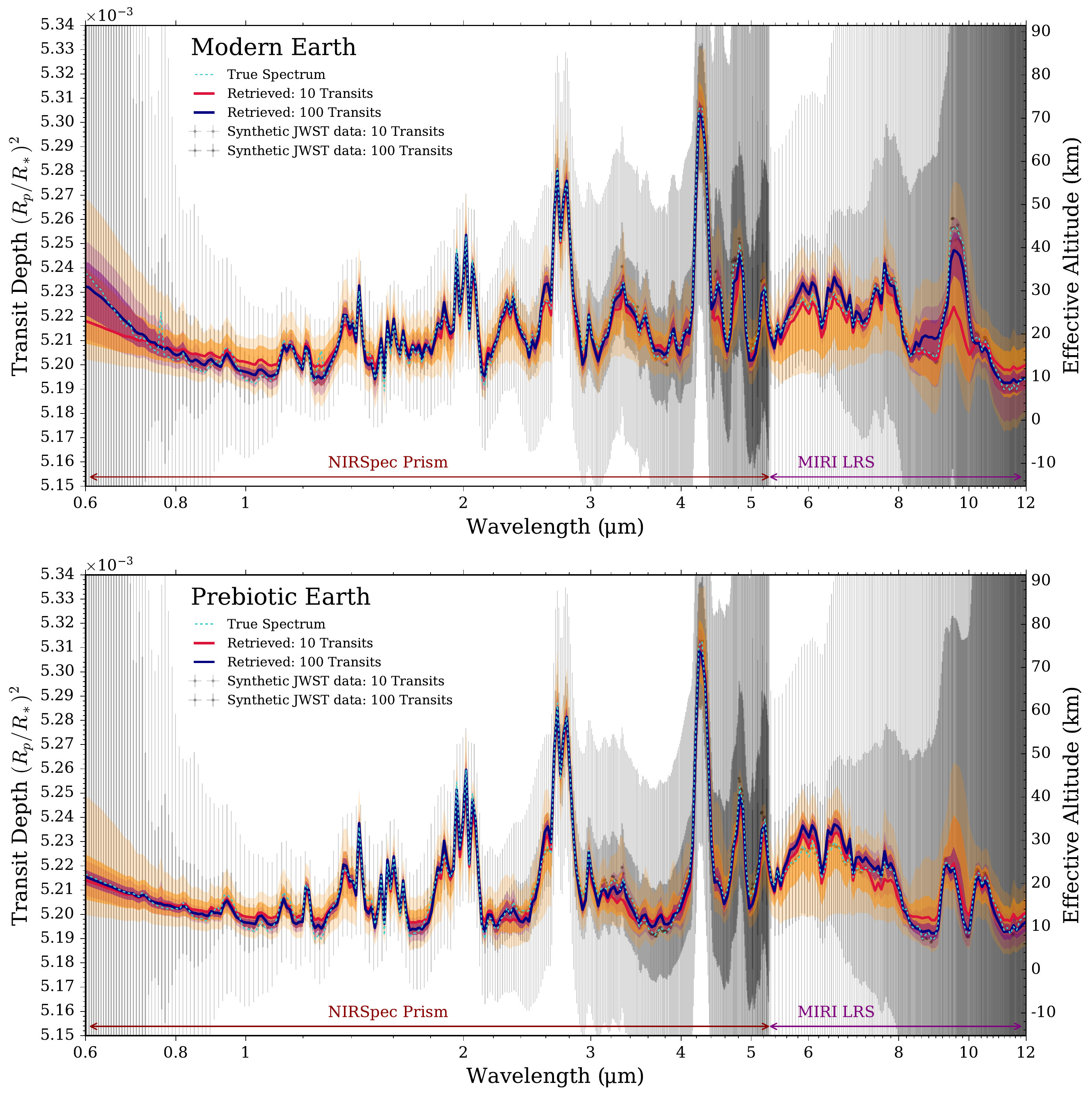}
    \caption{Retrieved transmission spectra from simulated \textit{JWST} observations of TRAPPIST-1e. Top: 1\,bar modern Earth scenario. Bottom: prebiotic Earth scenario. Simulated \textit{JWST} observations for 10 transits (light error bars) and 100 transits (dark error bars) with each of the NIRSpec Prism and MIRI LRS were subject to atmospheric retrievals by POSEIDON. The true model transmission spectra used to generate the simulated \textit{JWST} observations are overlaid (dashed blue curves), plotted at $R=100$ for clarity. The median retrieved transmission spectra and 1$\sigma$ / 2$\sigma$ confidence intervals are shown for two observing campaigns: 10 NIRSpec Prism + 10 MIRI LRS transits (orange shading) and 100 NIRSpec Prism + 100 MIRI LRS transits (purple shading). Similar confidence intervals are obtained with only NIRSpec Prism transits, as the larger MIRI LRS error bars add comparatively little information content. The prebiotic scenario is generally well-retrieved within the remit of a 10 Prism transit campaign. The defining spectral features of the modern Earth scenario, N$_2$O and O$_3$ signatures, require a more intensive 100 Prism transit campaign to be correctly identified.}
    \label{fig:retrieved_spectrum}
\end{figure*}

Our simulated \textit{JWST} datasets are shown in Figure~\ref{fig:retrieved_spectrum}. We opt to remove Gaussian scatter from the simulated observations for our retrievals, centering each datum on its (true) binned model location. \citet{Feng2018} showed that this results in posteriors representing the average over an ensemble of noise instances. We follow this approach to ensure our results are not specific to a given random noise draw. We verified the validity of this choice by running additional retrievals with Gaussian noise (as shown in Figure~\ref{fig:retrieved_spectra_Gauss_scatter}) finding consistent results to those described below. With our simulated \textit{JWST} observations in hand, we now describe our atmospheric retrieval approach.

\subsection{Atmospheric Retrieval Analysis} \label{subsec:retrieval_methods}

We explore the efficacy of extracting atmospheric properties from \textit{JWST} transmission spectra of TRAPPIST-1e using the POSEIDON atmospheric retrieval code \citep{MacDonald2017}. POSEIDON couples a parametric planetary atmosphere and radiative transfer model to a Bayesian parameter estimation and model comparison suite built on MultiNest \citep{Feroz2008,Feroz2009,Feroz2019} via the PyMultiNest package \citep{Buchner2014}. The adaptation of POSEIDON to model and retrieve transmission spectra of terrestrial planets is described in \citet{Kaltenegger2020}, here we summarise the salient aspects concerning our TRAPPIST-1e analysis.

We first generated `ground-truth' model transmission spectra for our modern and prebiotic Earth TRAPPIST-1e scenarios. At the low spectral resolution of \textit{JWST} NIRSpec Prism and MIRI LRS observations ($R \approx 50-400$), only a subset of the molecular species included in the high-resolution models presented in section\,\ref{sec:high_res_spectra} are observable. We therefore generated two new forward model transmission spectra for the 1 bar modern Earth and prebiotic Earth considering only the most spectrally prominent molecules: N$_2$, O$_2$, O$_3$, H$_2$O, CH$_4$, CO$_2$, and N$_2$O (see Figure \ref{fig:molecular_signatures}). The molecular cross sections used by POSEIDON for temperate planets are derived from HITRAN 2016 line lists \citep{Gordon2017}, with the addition of optical O$_3$ absorption \citep{Serdyuchenko2014} and the latest HITRAN CIA data \citep{Karman2019}. The altitude-dependent mixing ratio profiles for these molecules and the atmospheric temperature structure were interpolated from the models in section\,\ref{sec:atmospheric_models} onto an 81-layer vertical grid spaced uniformly in log-pressure from $10^{-7}$--$10\,$bar. To consider the influence of clouds on our results, we included an opaque cloud deck at 6 km altitude -- roughly commensurate with the middle cloud layer altitudes on Earth. We generated transmission spectra at $R =$ 10,000 from 0.4--14 $\micron$ for the modern and prebiotic Earth scenarios, serving as the input to our PandExo \textit{JWST} simulations. We note that the choice to generate new spectra forward models for our retrieval analysis, rather than using the high-resolution models in section~\ref{sec:high_res_spectra} was taken to ensure consistency in PandExo, which generates transmission spectra for retrievals and simulated \textit{JWST} observations from atmospheric models. This ensures that any differences between our retrieved atmospheric properties and the \textit{ground-truth} spectrum are due to assumptions in how the retrieval code parametrises the atmosphere (e.g. uniform mixing ratios), rather than biases from subtle differences between Exo-Prime2 and POSEIDON (e.g. line lists or CIA sources).

Atmospheric retrievals involve repeated calls to a radiative transfer model by a Bayesian sampling algorithm, which thereby identifies the range of parametrised atmospheric properties consistent with a dataset. In the present analysis, the pressure-temperature (P-T) profile is parametrised by an adaptation of the six-parameter function from \citet{Madhusudhan2009}, the mixing ratios of O$_2$, O$_3$, H$_2$O, CH$_4$, CO$_2$, and N$_2$O are each ascribed a single parameter (assumed uniform with altitude), and two further parameters describe the 1-bar planetary radius and cloud-top / surface pressure -- for a total of up to 14 free parameters. The N$_2$ mixing ratio is specified by the mixing ratio summation to unity condition. The priors for each parameter are uniform or uniform in the logarithm, with the following ranges: surface temperature (uniform: $100$--$400\,$K), mixing ratios (log-uniform: $10^{-12}$--$1$, subject to sums exceeding unity rejected), planetary radius (uniform: $0.8$--$1.2\,R_{\earth}$), and cloud pressure (log-uniform: $10^{-7}$--$10\,$bar). Retrievals only including Prism data compute transmission spectra at $R =$ 10,000 from $0.58$--$5.4\,\micron$, while those including MIRI LRS data extend from $0.58$--$12.1\,\micron$. Each retrieval uses 2,000 MultiNest live points during the parameter space exploration.

Our retrieval analysis spans a total of 66 retrievals covering each simulated \textit{JWST} campaign and both the modern Earth and prebiotic Earth scenarios. For each of the six observing campaigns (number of Prism or MIRI transits, see section~\ref{subsec:simulated_data}), we first ran one reference retrieval using all 14 free parameters described above. These reference retrievals are used for Bayesian parameter estimation (e.g. the predicted mixing ratio constraints in section~\ref{subsubsec:retrieved_abundances} and Figure~\ref{fig:retrieved_abundances}). Additional retrievals were run with one or more molecules excluded in the following combinations: no CO$_2$, no H$_2$O, no CH$_4$, no N$_2$O (modern only), no O$_3$ (modern only), and no N$_2$O + O$_3$ (modern only). The computed Bayesian evidence lowers when a molecule that influences the true spectrum is excluded from the fitting process - analogous to an increase in the reduced chi-square when a necessary model parameter is removed \citep{Trotta2017}. By comparing the Bayesian evidences of the reference model with the models lacking each molecule, we compute their predicted detection significances \citep[see e.g.][]{Benneke2013,MacDonald2017}.

\subsection{Differentiating Modern and Prebiotic Earth Scenarios for TRAPPIST-1e with \textit{JWST}} \label{subsec:retrieval_results}

Here we present the results of our atmospheric retrievals. In turn, we cover the ability to quantitatively recover the spectral features, P-T profiles, and chemical abundances from \textit{JWST} observations of TRAPPIST-1e. 

\subsubsection{Retrievability of Spectral Features} \label{subsubsec:retrieved_spectra}

We first provide a visual demonstration of which spectral features can be correctly recovered from \textit{JWST} observations of our TRAPPIST-1e model atmospheres. We focus here on the spectral information one can correctly infer from two campaigns: 10 NIRSpec Prism + 10 MIRI LRS transits vs. 100 NIRSpec Prism + 100 MIRI LRS transits (hereafter 10+10 transit and 100+100 transit). Figure~\ref{fig:retrieved_spectrum} compares the median retrieved transmission spectra from these two campaigns for both the modern and prebiotic Earth scenarios. 

The modern and prebiotic Earth scenarios require differing amounts of observing time to correctly retrieve their spectral features. For the prebiotic Earth, the spectral morphology of the input model -- driven mainly by CO$_2$ and H$_2$O -- is largely correctly retrieved with the 10+10 transit campaign. For the modern Earth, however, the 10+10 transit campaign results in large deviations between the input and retrieved model. This is caused by the large data uncertainties at the shortest Prism wavelengths and the longest MIRI wavelengths, preventing clear identification of O$_3$ (via the Chappuis and Wulf bands from 0.6 to $0.9\,\micron$ covered by the Prism and the $9.6\,\micron$ feature covered by MIRI). The larger 100+100 transit campaign successfully recovers these O$_3$ features, alongside the relatively weak N$_2$O feature around $4.6\,\micron$. The O$_2$ $A$-band ($0.76\,\micron$) is too weak to be identified by either campaign. We note that some residual differences between the input models and the retrieved spectra (e.g. for the H$_2$O near $6\,\micron$) likely arise from the assumption of uniform vertical abundance profiles in the retrievals \citep{Changeat2019}. We surmise from Figure~\ref{fig:retrieved_spectrum} that the 10+10 transit campaign can broadly recover the spectral signatures of the prebiotic Earth, but the larger 100+100 transit campaign can identify the distinguishing features of the modern Earth scenario. As we show below, this information content derives from the NIRSpec Prism data, with MIRI LRS transits largely unnecessary.

We quantify this intuitive picture via predicted detection significances for each molecule. Table~\ref{table:JWST_detection_significances} presents the resulting significances from our Bayesian model comparisons for the \textit{JWST} campaigns outlined in section~\ref{subsec:simulated_data}. These quoted significances encode the information content across the full wavelength range of the data, automatically factoring in any model degeneracies (e.g. overlapping molecular absorption features) into the Bayesian evidence calculations \citep[e.g.][]{Benneke2013,Trotta2017}. 

For the modern Earth scenario, molecular species can be detected at high confidence starting from 10 transits. We find 10 Prism transits are sufficient to detect CO$_2$ to $>$ $5\,\sigma$, but provide only weak evidence for H$_2$O and CH$_4$ ($1.8\,\sigma$ and $2.5\,\sigma$, respectively) -- partly due to the 6 km cloud deck obscuring layers near the surface, which contain most of the H$_2$O in our model. An increase to 20 Prism transits allows a $3.7\,\sigma$ detection of CH$_4$, but 100 Prism transits are required before CH$_4$ and H$_2$O can be independently detected to $>$ $5\,\sigma$. The low predicted significances for H$_2$O arise from its low abundance at the altitudes mainly probed by transmission spectra ($\sim 10^{-5}$ for $z >$ 10\,km - see Figure~\ref{fig:atm_profiles}). The detection of N$_2$O or O$_3$ will be challenging: while a 200 transit Prism campaign can detect N$_2$O to $3.7\,\sigma$, in none of our campaigns is O$_3$ detected to $>$ $3\,\sigma$. Similarly, O$_2$ is undetected in all our retrievals. A somewhat higher significance can result by searching for the \emph{agnostic combination} of N$_2$O + O$_3$, reaching $4.1\,\sigma$ for 200 Prism transits. 

For the prebiotic Earth a similar picture arises. Detections of CO$_2$ and H$_2$O have slightly higher predicted significances due to their enhanced prebiotic abundances compared to the modern Earth. However, detecting CH$_4$ is significantly more challenging for the prebiotic Earth -- requiring 200 Prism transits for a $4.2\,\sigma$ detection -- due to the prebiotic CH$_4$ abundance being lower than the modern Earth scenario by over an order of magnitude.

\newcommand{\ra}[1]{\renewcommand{\arraystretch}{#1}}
\begin{table}
\ra{1.2}
	\caption{Predicted molecular detection significances}
	\begin{tabular*}{\columnwidth}{l@{\extracolsep{\fill}} cccccccl@{}}\toprule
		Campaign & CO$_2$ & H$_2$O & CH$_4$ & N$_2$O & O$_3$ & \hspace{-10pt} N$_2$O + O$_3$  \\ \midrule
		\textbf{Modern} & \\ 
		10 Prism & $5.2\,\sigma$ & $1.8\,\sigma$ & $2.5\,\sigma$ & N/A & N/A & N/A \\
		20 Prism & $7.1\,\sigma$ & $2.4\,\sigma$ & $3.7\,\sigma$ & $1.5\,\sigma$ & N/A & $1.2\,\sigma$ \\
		100 Prism & $15.2\,\sigma$ & $5.4\,\sigma$ & $8.7\,\sigma$ & $2.6\,\sigma$ & $1.6\,\sigma$ & $2.7\,\sigma$ \\
		200 Prism & $21.4\,\sigma$ & $7.7\,\sigma$ & $12.2\,\sigma$ & $3.7\,\sigma$ & $2.4\,\sigma$ & $4.1\,\sigma$ \\
		10 Prism + & -- & -- & -- & -- & -- & -- \\
		10 MIRI & $5.2\,\sigma$ & $2.0\,\sigma$ & $2.4\,\sigma$ & N/A & N/A & N/A \\
		100 Prism + & -- & -- & -- & -- & -- & -- \\
		100 MIRI & $15.6\,\sigma$ & $6.7\,\sigma$ & $8.7\,\sigma$ & $2.5\,\sigma$ & $2.8\,\sigma$ & $3.5\,\sigma$ \\ \midrule
		\textbf{Prebiotic} & \\
		10 Prism & $6.0\,\sigma$ & $2.2\,\sigma$ & N/A & N/A & N/A & N/A \\
		20 Prism & $8.2\,\sigma$ & $3.0\,\sigma$ & N/A & N/A & N/A & N/A \\
		100 Prism & $17.7\,\sigma$ & $7.2\,\sigma$ & $2.6\,\sigma$ & N/A & N/A & N/A \\
		200 Prism & $25.0\,\sigma$ & $10.5\,\sigma$ & $4.2\,\sigma$ & N/A & N/A & N/A \\
		10 Prism + & -- & -- & -- & -- & -- & -- \\
		10 MIRI & $6.0\,\sigma$ & $2.5\,\sigma$ & N/A & N/A & N/A & N/A \\
		100 Prism + & -- & -- & -- & -- & -- & -- \\
		100 MIRI & $18.3\,\sigma$ & $8.5\,\sigma$ & $2.7\,\sigma$ & N/A & N/A & N/A \\
		\bottomrule
		\vspace{0.1pt}
	\end{tabular*}
	$\textbf{Notes}:$ `Prism' refers to \textit{JWST} NIRSpec Prism transit observations, while `MIRI' refers to \textit{JWST} MIRI LRS. `N/A' indicates our retrieval Bayesian model comparisons yield no evidence (Bayes factor $\lesssim$ 1) for the presence of a given molecule.
	\label{table:JWST_detection_significances}
\end{table}

\begin{figure*}
	\includegraphics[width=\textwidth, trim={0.0cm 0.3cm 0.0cm 0.0cm}]{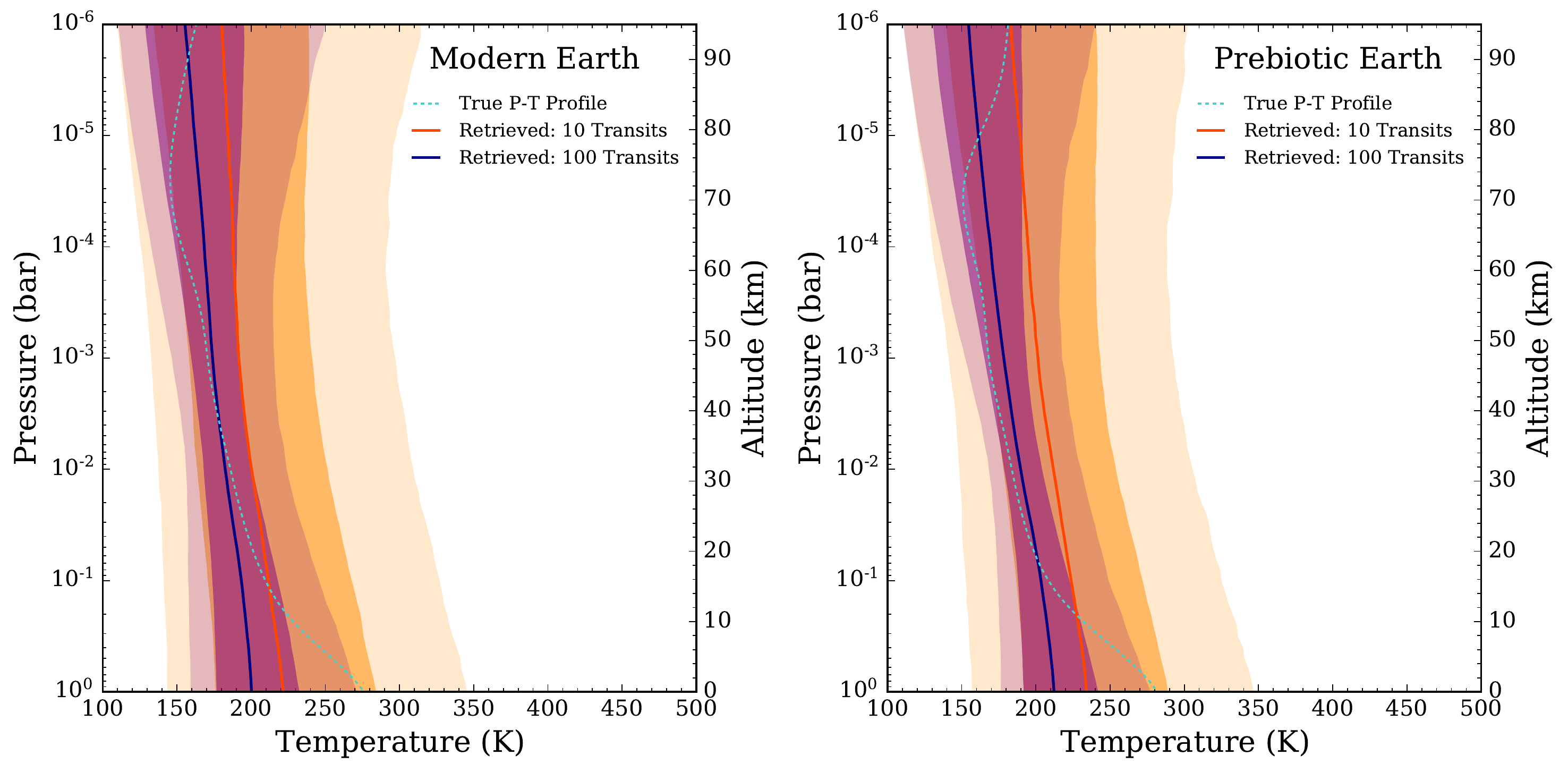}
    \caption{Retrieved pressure-temperature structures from simulated \textit{JWST} observations of TRAPPIST-1e. Left: 1\,bar modern Earth scenario. Right: prebiotic Earth scenario. The true P-T profiles for each scenario are overlaid (dashed blue curves). The median retrieved P-T profiles and 1$\sigma$ / 2$\sigma$ confidence intervals are shown for two observing campaigns: 10 NIRSpec Prism transits (orange shading) and 100 NIRSpec Prism transits (purple shading). Both campaigns correctly constrain the stratospheric temperature, reaching $\sim$ 30-50\,K precision for the 10 Prism transit campaign and $\sim$ 10-20\,K for the 100 Prism transit campaign. The surface temperature is not reliably retrieved, due to the lack of spectral sensitivity to pressures $> 0.1$\,bar and the cloud deck at 6\,km altitude.}
    \label{fig:retrieved_PT}
\end{figure*}

We conclude from the above that a minimum of 100 Prism transits are required to identify evidence differentiating our modern and prebiotic scenarios solely from spectral feature detections / non-detections (e.g. N$_2$O + O$_3$). However, as discussed below, the retrieved abundance constraints offer a second line of evidence to enable their distinguishability.  

Our results show that transmission spectra observations of TRAPPIST-1e with the NIRSpec Prism generally yield more atmospheric information than MIRI LRS observations. Our retrievals including MIRI data result in improved O$_3$ and H$_2$O detection significances. However, an equal time with additional Prism observations (e.g. 200 Prism vs. 100 Prism + 100 MIRI) results in much higher detection significances for H$_2$O, CO$_2$, CH$_4$, and N$_2$O. Even 100 transits with MIRI only slightly improves the significance of O$_3$ ($2.8\,\sigma$ vs. $2.4\,\sigma$). Since the Prism covers both N$_2$O and O$_3$ features, evidence for these important biosignature components is obtained more efficiently by stacking additional Prism transits without the need for Prism + MIRI campaigns.

\subsubsection{Constraining Pressure-Temperature Profiles} \label{subsubsec:retrieved_PT}

Temperature structure constraints derived from \textit{JWST} observations will aid our understanding of the habitability of TRAPPIST-1e. Figure~\ref{fig:retrieved_PT} shows the retrieved P-T profiles from the 10 Prism transit and 100 Prism transit campaigns. For both the modern and prebiotic Earth scenarios, the stratospheric temperature is generally correctly retrieved within the $1\,\sigma$ confidence region. Temperatures are recovered to $\sim 30$--$50$\,K accuracy for the 10 Prism transit campaign, with an improvement to $\sim 10$--$20$\,K for the 100 Prism transit campaign. The surface temperature is not reliably retrieved in either scenario, as our transmission spectra are only weakly sensitive to pressures $>$ 0.1 bar (partly due to the 6 km cloud deck obscuring layers near the surface). Since the P-T profiles for the modern and prebiotic Earth scenarios are broadly similar, temperature profile constraints do not offer an avenue to distinguish these models. However, it is encouraging that the first-order shape of the P-T profiles (i.e. the stratospheric temperature scale and its gradient with pressure) can be constrained within even a 10 Prism transit \textit{JWST} campaign.  

\subsubsection{Measuring the Atmospheric Composition of TRAPPIST-1e with \textit{JWST}} \label{subsubsec:retrieved_abundances}

\begin{figure*}
	\includegraphics[width=0.92\textwidth]{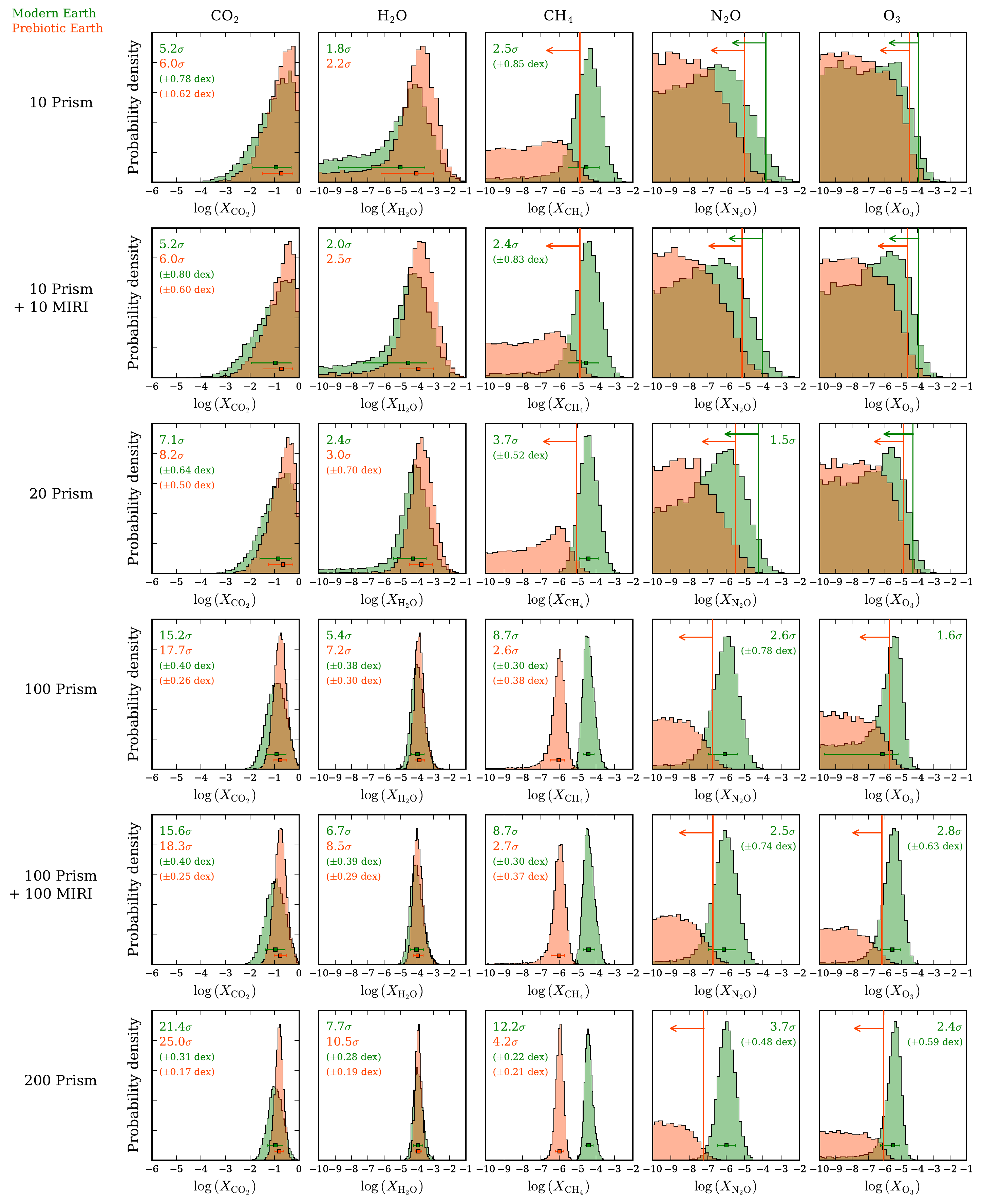}
    \caption{Retrieved atmospheric composition from simulated \textit{JWST} observations of TRAPPIST-1e. The predicted mixing ratio constraints for CO$_2$, H$_2$O, CH$_4$, N$_2$O, and O$_3$ for the 1\,bar modern Earth (green) and prebiotic Earth (orange) are shown by the marginalised posterior histograms. Each row corresponds to a different \textit{JWST} observing campaign, ranging from 10 NIRSpec Prism transits (top) to 200 Prism transits (bottom). Various combinations of NIRSpec Prism + MIRI LRS transits are also included. Predicted detection significances for each molecule are annotated (green text for the modern Earth; orange text for the prebiotic Earth). Abundance constraints (error bars) are overlaid, with the $\pm 1\,\sigma$ interval annotated where constraints $<$ 1\,dex are possible. For non-detections, $2\,\sigma$ abundance upper limits are shown (vertical lines with arrows). The prebotic and Modern scenarios can first be distinguished with 20 Prism transits via the CH$_4$ abundance (the $2\,\sigma$ prebiotic upper limit precludes the modern abundance). Similarly, the N$_2$O abundance can be differentiated with 200 Prism transits. Even 200 Prism transits (or 100 Prism + 100 MIRI transits) are insufficient to distinguish the O$_3$ abundances. O$_2$ is unconstrained in all cases. Adding MIRI data slightly increases the detection significances of O$_3$, but adding the same quantity of Prism transits produces tighter constraints on all other atmospheric parameters.}
    \label{fig:retrieved_abundances}
\end{figure*}

Molecular abundance constraints offer another avenue to differentiate modern and prebiotic scenarios. Even when a molecule is undetected in one scenario, an upper limit on its abundance can suffice to disfavour biological production for certain molecules. Figure~\ref{fig:retrieved_abundances} shows the posterior distributions for the mixing ratio of each molecule from our retrievals. A 10 Prism transit campaign suffices to constrain the CO$_2$ and CH$_4$ abundances of the modern Earth to within an order of magnitude (0.78 dex and 0.85 dex, respectively); for the prebiotic Earth only CO$_2$ is constrained (0.62 dex). An increase to 20 Prism transits offers two new results: (i) an order of magnitude constraint on the H$_2$O abundance for the prebiotic Earth (0.70 dex); and (ii) a $2\,\sigma$ upper limit on the prebiotic Earth CH$_4$ abundance in tension with the $1\,\sigma$ confidence interval of the modern Earth CH$_4$ abundance. The latter is a diagnostic difference between the modern and prebiotic scenarios accessible with only 20 transits. A further increase to 100 Prism transits offers dramatic improvements: constraints on the CO$_2$, H$_2$O, and CH$_4$ abundances to $\sim$ a factor of 2 (0.3 dex), and an order of magnitude constraint on the modern Earth N$_2$O abundance (0.78 dex). Constraints on the O$_3$ abundance of the modern Earth can be obtained, to within an order of magnitude, with either 100 Prism + 100 MIRI transits (0.63 dex), or 200 Prism transits (0.59 dex). However, 200 Prism transits result in significantly better constrained abundances for the other molecules. As above, we recommend additional Prism transits instead of dividing time between the Prism and MIRI.

Our analysis demonstrates that the most effective way to differentiate our modern and prebiotic Earth scenarios is via differences in their retrieved CH$_4$ abundances. A 20 Prism transit campaign would provide evidence supporting either the modern scenario (via a constrained abundance) or the prebiotic scenario (via an upper limit). This tentative evidence can then be confirmed by additional followup observations, with a 100-transit campaign conclusively differentiating the CH$_4$ abundances for each scenario. While we find we can differentiate the CH$_4$ abundances for our modern and prebiotic scenarios, the true CH$_4$ abundance of TRAPPIST-1e could span a wide continuum. Therefore, any constrained CH$_4$ abundance must be carefully considered in context before attribution to a biological flux.

N$_2$O abundance measurements offer a secondary, albeit observationally intensive, manner to differentiate our scenarios. A 100-transit Prism campaign offers a bounded constraint on the N$_2$O abundance for our modern scenario (see Figure~\ref{fig:retrieved_abundances}). An increase to 200 Prism transits would secure the N$_2$O detection, while placing an upper limit on the N$_2$O abundance for the prebiotic scenario outside the confidence interval for the modern Earth N$_2$O abundance. However, we note that 200 transits are likely beyond the capabilities of feasible campaigns with \textit{JWST}. An estimated 81 transits of TRAPPIST-1e will occur over \textit{JWST}'s nominal 5-year mission \citep{Lustig-Yaeger2019}, so our simulated observations exceeding 100 transits can be considered a theoretical exercise or a best case scenario for an extended \textit{JWST} mission. Nevertheless, we stress that molecular abundance determinations, may offer a promising path to differentiate modern and prebiotic scenarios for TRAPPIST-1e.

\section{Summary \& Discussion} \label{sec:summary_discussion}

TRAPPIST-1e offers the prospect of detailed atmospheric profiling for a terrestrial exoplanet in the near future. In this study, we considered avenues to differentiate between modern and prebiotic Earth scenarios for TRAPPIST-1e via transmission spectroscopy. We used the new Mega-MUSCLES SED for TRAPPIST-1 to generate self-consistent atmospheres and high-resolution transmission spectra ($R \sim 100,000$) for two modern Earth scenarios (with 1\,bar and 0.5\,bar surface pressure, both including biotic flux) and a 1\,bar prebiotic Earth scenario. Our high-resolution model spectra are available \href{http://doi.org/10.5281/zenodo.4770258}{online} to optimize and interpret upcoming observations. We also presented an atmospheric retrieval analysis, demonstrating that \textit{JWST} is capable of differentiating between modern and prebiotic Earth scenarios. Our main results are as follows:

\begin{enumerate}
    \item At high-resolution, prominent O$_2$ and O$_3$ features, combined with enhanced CH$_4$ features, are a strong diagnostic for biotic flux in modern Earth scenarios.
    \item With \textit{JWST}, the most time-effective way to differentiate our modern and prebiotic Earth scenarios is via CH$_4$ abundance measurements. As little as 20 NIRSPec Prism transits can distinguish modern from prebiotic CH$_4$ abundances for our two specific models.
    \item Precise molecular abundance measurements are possible. A 10-transit Prism program can constrain CO$_2$ and CH$_4$ to within an order of magnitude for our modern Earth scenario, while CO$_2$ alone can be constrained for our prebiotic Earth scenario. A moderate 20-transit Prism program can measure CO$_2$, CH$_4$, or H$_2$O abundances to a precision of $\lesssim$ 0.7 dex (a factor of 5). Tighter abundance constraints are obtained by focusing on Prism transits instead of splitting observing resources between Prism and MIRI LRS transits.
    \item Detections of H$_2$O are challenging (due to low stratospheric abundances), requiring $\sim$ 100 transits to reach 5\,$\sigma$.
    \item Evidence of O$_3$ will be extremely difficult to observe for our modern Earth scenario with \textit{JWST}, only reaching 1.6\,$\sigma$ with 100 Prism transits. MIRI transits can aid O$_3$ detectability, but even 100 Prism + 100 MIRI transits (beyond the nominal lifetime of \textit{JWST}) would just reach 2.8\,$\sigma$.
    \item Evidence of N$_2$O, an important component of biosignatures, can emerge within 100 Prism transits at 2.6\,$\sigma$ confidence. Near-infrared spectral signatures of N$_2$O are more readily identified with \textit{JWST} than O$_3$.
    \item Stratospheric temperatures can be measured to $\sim 30$--$50$\,K precision with 10 Prism transits and $\sim 10$--$20$\,K with 100 Prism transits.
\end{enumerate}

\noindent Our findings have several implications, discussed below.

\subsection{The Challenge of Detecting Oxygen-Based Biosignatures with \textit{JWST}} \label{subsec:biosignature_challenges}

The canonical biosignature features an oxidising gas in combination with a reducing gas, such that both species require a continuous supply: In the context of Earth, O$_2$ + CH$_4$ or O$_2$ + N$_2$O are generally considered \citep{Lovelock_1965Natur.207..568L,Lederberg_1965Natur.207....9L}. In oxygen-rich environments, O$_3$ is often taken as a proxy for O$_2$ \citep{Kasting_2014PNAS..11112641K,kaltenegger_how_2017,Schwieterman_2018AsBio..18..663S}.

Our results demonstrate that high-confidence detections of oxygen-based biosignatures will be extremely challenging with \textit{JWST}. The reducing component of the O$_2$ + CH$_4$ biosignature pair, CH$_4$, can be strongly detected ($> 3.6\,\sigma$) with 20 transits for our modern Earth scenario. However, detections of oxidizing gases will prove considerably more time-intensive. This arises because the most prominent O$_2$ and O$_3$ features lie at wavelengths where both the \textit{JWST} NIRSPec Prism and MIRI LRS have relatively little sensitivity (see Figure~\ref{fig:retrieved_spectrum}). Consequently, we find that even 200 Prism transits are insufficient to detect O$_2$. Hints of O$_3$ signatures can be discerned within 100 transits, but fall short of 2\,$\sigma$ confidence even for such a large program. We suggest that stronger evidence of modern Earth-like biosignatures can be obtained via the agnostic combination of N$_2$O and/or O$_3$, which our retrievals predict can be detected at 2.7\,$\sigma$ with 100 Prism transits. If such a detection arises from O$_3$, one has the canonical O$_3$ + CH$_4$ biosignature pair (since CH$_4$ is detectable within 20 transits). If the detection instead arises from N$_2$O, this could indicate a large N$_2$O flux arising from denitrifying bacteria and hence constitute a biosignature \citep[see e.g.][]{Seager2012}.

The large time commitment ($\gtrsim$ 100 transits) required to detect biosignatures in TRAPPIST-1e's atmosphere with \textit{JWST} requires careful consideration of limited telescope resources \citep{Gillon_2020arXiv200204798G}. Currently, 4 transits of TRAPPIST-1e will be observed by the NIRSpec Prism for the \textit{JWST} GTO, with no scheduled Cycle 1 General Observer (GO) programs. We have shown that, under the assumption of a modern Earth-like CH$_4$ production flux, hints of CH$_4$ can emerge with 20 NIRSpec Prism transits. To achieve this, a program that at least quadruples the scheduled number of transits would be required. Our retrieval analysis finds that Prism transits alone generally place better constrains on atmospheric properties, such as molecular abundances, compared with NIRSpec Prism + MIRI LRS observations with the same total number of transits. We therefore recommend that the community focus on building a large Prism program for TRAPPIST-1e, rather than dividing time between the Prism and MIRI.

We additionally expect that non-grey scattering from aerosols, not elaborated in this paper, will likely complicate biosignature detectability. At visible wavelengths, the detection of O$_3$ relies on the Chappuis band at 0.6 $\micron$. However, scattering from hazes can produce strong slopes over this wavelength range \citep[e.g.,][]{Marley2013,Ohno2020} and potentially obscure O$_3$. In Figure~\ref{fig:highres_features} (first column), we show the high-resolution O$_3$ Chappuis band compared to the Rayleigh scattering slope. Haze absorption can have a similar morphology to the Rayleigh slope (i.e. a slope, perhaps with a different gradient, but without distinct spectral features). The O$_3$ Chappuis feature, however, has a unique band structure that can be distinguished from a Rayleigh or haze slope, providing any haze extinction is not strong enough to obscure the feature entirely. Even if a strong haze is present, a low CH$_4$/CO$_2$ ratio could imply that such hazes have a biological origin \citep{Arney_2018AsBio..18..311A}. Given that we find 10 JWST transits can constrain CH$_4$ and CO$_2$ abundances to within an order of magnitude for our modern Earth model, a detection of haze could serve as an alternative hint of biological activity within a medium \textit{JWST} program. Attribution to biology would however require careful modelling of haze production mechanisms.

The spatial distribution and temporal variability of clouds can also complicate the detection of spectral features. Our models conservatively assume an optically thick cloud deck uniformly covering the day-night terminator of TRAPPIST-1e with an altitude of 6 km, corresponding to the middle cloud layer on Earth. However, the tidally locked nature of TRAPPIST-1e can lead to clouds with non-uniform 3D properties that can alter the strength of spectral features compared to 1D models \citep[e.g.][]{fauchez_impact_2019}. One additionally expects that temporal cloud variability will substantially influence on the planet's climate by controlling the radiative balance. In the context TRAPPIST-1e, however, cloudy variability is unlikely to affect retrieved abundances when multiple transits are stacked \citep{May2021}.

\subsection{Opportunities for High-Resolution Observations} \label{subsec:high-res_opportunities}

At high spectral resolution ($R \sim 100,000$), signatures of O$_2$ and O$_3$ display much greater feature contrasts than can be accessed with \textit{JWST} ($R < 3,000$). Our high-resolution models display clear differences between the modern and prebiotic Earth scenarios via O$_2$, O$_3$, and CH$_4$ lines (Figure~\ref{fig:highres_features}), holding the promise of differentiating these scenarios with upcoming high-resolution spectrographs with the new generation of ELTs \citep[e.g.][]{Snellen2013,Rodler2014,Serindag2019,Hawker2019}. 
Upcoming spectrographs like HIRES ($\sim 0.3$--$2.4\,\micron$) and METIS ($\sim 3$--$19\,\micron$) on the ELT are designed for a spectral resolutions of $100,000$--$150,000$ to distinguish Earth’s atmospheric features from those of potentially habitable exoplanets. The O$_3$ Chappuis band, the O$_2$ \textit{A}-band, and many CH$_4$ features fall within this wavelength range, where the photometric \textit{r'}-, \textit{i'}-, \textit{Z}-, \textit{Y}-, \textit{J}-, and \textit{H}-band provide transparent atmospheric windows. We note that the O$_3$ Chappuis band is a largely continuous band, even at high resolution \citep{Serdyuchenko2014}, but its distinctive shape compared to scattering slopes may allow its detection. We provide our high-resolution model transmission spectra from the visible to the infrared online to aid future theoretical and observational studies of TRAPPIST-1e.

\subsection{The Future of Biosignature Searches in TRAPPIST-1e's Atmosphere} \label{subsec:future_telescopes}

Although detecting weak evidence of biosignature pairs, such as O$_3$ + CH$_4$, in TRAPPIST-1e's atmosphere is possible with \textit{JWST}, any such inferences will be rather inconclusive ($< 3\,\sigma$) and require a substantial time allocation ($\gtrsim$ 100 transits). Nevertheless, molecular abundance measurements of other key species -- such as H$_2$O, CO$_2$, CH$_4$, and N$_2$O -- promise that such an endeavour with \textit{JWST} will provide vital context on the potential habitability of TRAPPIST-1e. Ground-based ELTs can provide additional information to constrain the atmospheric composition of Trappist-1e.

The \textit{JWST} era will set the stage for future space missions -- such as \textit{Origins}, \textit{HabEx} and \textit{LUOVIR} -- designed to use larger collecting areas to place statistical constraints on the frequency of life in our solar neighborhood. TRAPPIST-1e remains a promising target in our search for life in the universe, with spectra of its atmosphere sure to surprise and dazzle us in equal measure.

\section*{Acknowledgements}

We thank the anonymous referee for a constructive report. Z.L. and L.K. acknowledge funding from the Carl Sagan Institute at Cornell University.

\section*{Data Availability}

The high-resolution spectra, P-T profiles, and mixing ratio profiles for all our models are publicly available online at \href{http://doi.org/10.5281/zenodo.4770258}{http://doi.org/10.5281/zenodo.4770258}. The full posterior distributions from our retrievals are available as supplementary online material and at the above link.

\bibliographystyle{mnras}
\bibliography{Trap1e_citation}

\begin{thebibliography}{}
\makeatletter
\relax
\def\mn@urlcharsother{\let\do\@makeother \do\$\do\&\do\#\do\^\do\_\do\%\do\~}
\def\mn@doi{\begingroup\mn@urlcharsother \@ifnextchar [ {\mn@doi@}
  {\mn@doi@[]}}
\def\mn@doi@[#1]#2{\def\@tempa{#1}\ifx\@tempa\@empty \href
  {http://dx.doi.org/#2} {doi:#2}\else \href {http://dx.doi.org/#2} {#1}\fi
  \endgroup}
\def\mn@eprint#1#2{\mn@eprint@#1:#2::\@nil}
\def\mn@eprint@arXiv#1{\href {http://arxiv.org/abs/#1} {{\tt arXiv:#1}}}
\def\mn@eprint@dblp#1{\href {http://dblp.uni-trier.de/rec/bibtex/#1.xml}
  {dblp:#1}}
\def\mn@eprint@#1:#2:#3:#4\@nil{\def\@tempa {#1}\def\@tempb {#2}\def\@tempc
  {#3}\ifx \@tempc \@empty \let \@tempc \@tempb \let \@tempb \@tempa \fi \ifx
  \@tempb \@empty \def\@tempb {arXiv}\fi \@ifundefined
  {mn@eprint@\@tempb}{\@tempb:\@tempc}{\expandafter \expandafter \csname
  mn@eprint@\@tempb\endcsname \expandafter{\@tempc}}}

\bibitem[\protect\citeauthoryear{{Agol} et~al.,}{{Agol}
  et~al.}{2021}]{Agol2021PSJ.....2....1A}
{Agol} E.,  et~al., 2021, \mn@doi [The Planetary Science Journal]
  {10.3847/PSJ/abd022}, \href
  {https://ui.adsabs.harvard.edu/abs/2021PSJ.....2....1A} {2, 1}

\bibitem[\protect\citeauthoryear{Airapetian, Glocer, Khazanov, Loyd, France,
  Sojka, Danchi  \& Liemohn}{Airapetian et~al.}{2017}]{airapetian_how_2017}
Airapetian V.~S.,  Glocer A.,  Khazanov G.~V.,  Loyd R. O.~P.,  France K.,
  Sojka J.,  Danchi W.~C.,   Liemohn M.~W.,  2017, \mn@doi [ApJ]
  {10.3847/2041-8213/836/1/L3}, 836, L3

\bibitem[\protect\citeauthoryear{{Allard}}{{Allard}}{2016}]{Allard_2016sf2a.conf..223A}
{Allard} F.,  2016, in {Reyl{\'e}} C.,  {Richard} J.,  {Cambr{\'e}sy} L.,
  {Deleuil} M.,  {P{\'e}contal} E.,  {Tresse} L.,   {Vauglin} I.,  eds,
  SF2A-2016: Proceedings of the Annual meeting of the French Society of
  Astronomy and Astrophysics. pp 223--227

\bibitem[\protect\citeauthoryear{{Arney}, {Domagal-Goldman}  \&
  {Meadows}}{{Arney} et~al.}{2018}]{Arney_2018AsBio..18..311A}
{Arney} G.,  {Domagal-Goldman} S.~D.,   {Meadows} V.~S.,  2018, \mn@doi
  [Astrobiology] {10.1089/ast.2017.1666}, \href
  {https://ui.adsabs.harvard.edu/abs/2018AsBio..18..311A} {18, 311}

\bibitem[\protect\citeauthoryear{{Bailer-Jones}, {Rybizki}, {Fouesneau},
  {Mantelet}  \& {Andrae}}{{Bailer-Jones}
  et~al.}{2018}]{Bailer-Jones_2018AJ....156...58B}
{Bailer-Jones} C.~A.~L.,  {Rybizki} J.,  {Fouesneau} M.,  {Mantelet} G.,
  {Andrae} R.,  2018, \mn@doi [\aj] {10.3847/1538-3881/aacb21}, \href
  {https://ui.adsabs.harvard.edu/abs/2018AJ....156...58B} {156, 58}

\bibitem[\protect\citeauthoryear{{Baraffe}, {Homeier}, {Allard}  \&
  {Chabrier}}{{Baraffe} et~al.}{2015}]{Baraffe_2015A&A...577A..42B}
{Baraffe} I.,  {Homeier} D.,  {Allard} F.,   {Chabrier} G.,  2015, \mn@doi
  [\aap] {10.1051/0004-6361/201425481}, \href
  {https://ui.adsabs.harvard.edu/abs/2015A&A...577A..42B} {577, A42}

\bibitem[\protect\citeauthoryear{{Barstow} \& {Irwin}}{{Barstow} \&
  {Irwin}}{2016}]{Barstow_2016MNRAS.461L..92B}
{Barstow} J.~K.,  {Irwin} P.~G.~J.,  2016, \mn@doi [\mnras]
  {10.1093/mnrasl/slw109}, \href
  {https://ui.adsabs.harvard.edu/abs/2016MNRAS.461L..92B} {461, L92}

\bibitem[\protect\citeauthoryear{{Batalha} et~al.,}{{Batalha}
  et~al.}{2017}]{Batalha2017}
{Batalha} N.~E.,  et~al., 2017, \mn@doi [\pasp] {10.1088/1538-3873/aa65b0},
  \href {https://ui.adsabs.harvard.edu/abs/2017PASP..129f4501B} {129, 064501}

\bibitem[\protect\citeauthoryear{{Batalha}, {Lewis}, {Line}, {Valenti}  \&
  {Stevenson}}{{Batalha} et~al.}{2018}]{Batalha2018}
{Batalha} N.~E.,  {Lewis} N.~K.,  {Line} M.~R.,  {Valenti} J.,   {Stevenson}
  K.,  2018, \mn@doi [\apjl] {10.3847/2041-8213/aab896}, \href
  {https://ui.adsabs.harvard.edu/abs/2018ApJ...856L..34B} {856, L34}

\bibitem[\protect\citeauthoryear{{Benneke} \& {Seager}}{{Benneke} \&
  {Seager}}{2013}]{Benneke2013}
{Benneke} B.,  {Seager} S.,  2013, \mn@doi [\apj]
  {10.1088/0004-637X/778/2/153}, \href
  {https://ui.adsabs.harvard.edu/abs/2013ApJ...778..153B} {778, 153}

\bibitem[\protect\citeauthoryear{{Benneke} et~al.,}{{Benneke}
  et~al.}{2019}]{Benneke2019}
{Benneke} B.,  et~al., 2019, \mn@doi [\apjl] {10.3847/2041-8213/ab59dc}, \href
  {https://ui.adsabs.harvard.edu/abs/2019ApJ...887L..14B} {887, L14}

\bibitem[\protect\citeauthoryear{{Berger}, {Huber}, {van Saders}, {Gaidos},
  {Tayar}  \& {Kraus}}{{Berger} et~al.}{2020}]{Berger_2020AJ....159..280B}
{Berger} T.~A.,  {Huber} D.,  {van Saders} J.~L.,  {Gaidos} E.,  {Tayar} J.,
  {Kraus} A.~L.,  2020, \mn@doi [\aj] {10.3847/1538-3881/159/6/280}, \href
  {https://ui.adsabs.harvard.edu/abs/2020AJ....159..280B} {159, 280}

\bibitem[\protect\citeauthoryear{Bryson, Coughlin, Batalha, Berger, Huber,
  Burke, Dotson  \& Mullally}{Bryson et~al.}{2020}]{Bryson2020}
Bryson S.,  Coughlin J.,  Batalha N.~M.,  Berger T.,  Huber D.,  Burke C.,
  Dotson J.,   Mullally S.~E.,  2020, \mn@doi [The Astronomical Journal]
  {10.3847/1538-3881/ab8a30}, 159, 279

\bibitem[\protect\citeauthoryear{{Buchner} et~al.,}{{Buchner}
  et~al.}{2014}]{Buchner2014}
{Buchner} J.,  et~al., 2014, \mn@doi [Astronomy and Astrophysics]
  {10.1051/0004-6361/201322971}, \href
  {https://ui.adsabs.harvard.edu/\#abs/2014Astronomy and
  Astrophysics...564A.125B} {564, A125}

\bibitem[\protect\citeauthoryear{Bétrémieux \& Kaltenegger}{Bétrémieux \&
  Kaltenegger}{2014}]{betremieux_impact_2014}
Bétrémieux Y.,  Kaltenegger L.,  2014, \mn@doi [ApJ]
  {10.1088/0004-637X/791/1/7}, 791, 7

\bibitem[\protect\citeauthoryear{{Changeat}, {Edwards}, {Waldmann}  \&
  {Tinetti}}{{Changeat} et~al.}{2019}]{Changeat2019}
{Changeat} Q.,  {Edwards} B.,  {Waldmann} I.~P.,   {Tinetti} G.,  2019, \mn@doi
  [\apj] {10.3847/1538-4357/ab4a14}, \href
  {https://ui.adsabs.harvard.edu/abs/2019ApJ...886...39C} {886, 39}

\bibitem[\protect\citeauthoryear{{Cutri} et~al.,}{{Cutri}
  et~al.}{2003}]{Cutri2003}
{Cutri} R.~M.,  et~al., 2003, {2MASS All Sky Catalog of point sources.}

\bibitem[\protect\citeauthoryear{Dong, Lingam, Ma  \& Cohen}{Dong
  et~al.}{2017}]{dong_is_2017}
Dong C.,  Lingam M.,  Ma Y.,   Cohen O.,  2017, \mn@doi [ApJ]
  {10.3847/2041-8213/aa6438}, 837, L26

\bibitem[\protect\citeauthoryear{Dressing \& Charbonneau}{Dressing \&
  Charbonneau}{2015}]{dressing_occurrence_2015}
Dressing C.~D.,  Charbonneau D.,  2015, \mn@doi [ApJ]
  {10.1088/0004-637X/807/1/45}, 807, 45

\bibitem[\protect\citeauthoryear{{Ducrot} et~al.,}{{Ducrot}
  et~al.}{2020}]{Ducrot2020}
{Ducrot} E.,  et~al., 2020, \mn@doi [\aap] {10.1051/0004-6361/201937392}, \href
  {https://ui.adsabs.harvard.edu/abs/2020A&A...640A.112D} {640, A112}

\bibitem[\protect\citeauthoryear{Fauchez et~al.,}{Fauchez
  et~al.}{2019}]{fauchez_impact_2019}
Fauchez T.~J.,  et~al., 2019, \mn@doi [ApJ] {10.3847/1538-4357/ab5862}, 887,
  194

\bibitem[\protect\citeauthoryear{{Fauchez} et~al.,}{{Fauchez}
  et~al.}{2020}]{Fauchez2020}
{Fauchez} T.~J.,  et~al., 2020, \mn@doi [Nature Astronomy]
  {10.1038/s41550-019-0977-7}, \href
  {https://ui.adsabs.harvard.edu/abs/2020NatAs...4..372F} {4, 372}

\bibitem[\protect\citeauthoryear{{Feng}, {Robinson}, {Fortney}, {Lupu},
  {Marley}, {Lewis}, {Macintosh}  \& {Line}}{{Feng} et~al.}{2018}]{Feng2018}
{Feng} Y.~K.,  {Robinson} T.~D.,  {Fortney} J.~J.,  {Lupu} R.~E.,  {Marley}
  M.~S.,  {Lewis} N.~K.,  {Macintosh} B.,   {Line} M.~R.,  2018, \mn@doi [\aj]
  {10.3847/1538-3881/aab95c}, \href
  {https://ui.adsabs.harvard.edu/abs/2018AJ....155..200F} {155, 200}

\bibitem[\protect\citeauthoryear{{Feroz} \& {Hobson}}{{Feroz} \&
  {Hobson}}{2008}]{Feroz2008}
{Feroz} F.,  {Hobson} M.~P.,  2008, \mn@doi [MNRAS]
  {10.1111/j.1365-2966.2007.12353.x}, \href
  {https://ui.adsabs.harvard.edu/\#abs/2008MNRAS.384..449F} {384, 449}

\bibitem[\protect\citeauthoryear{{Feroz}, {Hobson}  \& {Bridges}}{{Feroz}
  et~al.}{2009}]{Feroz2009}
{Feroz} F.,  {Hobson} M.~P.,   {Bridges} M.,  2009, \mn@doi [MNRAS]
  {10.1111/j.1365-2966.2009.14548.x}, \href
  {https://ui.adsabs.harvard.edu/abs/2009MNRAS.398.1601F} {398, 1601}

\bibitem[\protect\citeauthoryear{{Feroz}, {Hobson}, {Cameron}  \&
  {Pettitt}}{{Feroz} et~al.}{2019}]{Feroz2019}
{Feroz} F.,  {Hobson} M.~P.,  {Cameron} E.,   {Pettitt} A.~N.,  2019, \mn@doi
  [The Open Journal of Astrophysics] {10.21105/astro.1306.2144}, \href
  {https://ui.adsabs.harvard.edu/abs/2019OJAp....2E..10F} {2, 10}

\bibitem[\protect\citeauthoryear{France et~al.,}{France
  et~al.}{2016}]{france_muscles_2016}
France K.,  et~al., 2016, \mn@doi [ApJ] {10.3847/0004-637X/820/2/89}, 820, 89

\bibitem[\protect\citeauthoryear{Fujii et~al.,}{Fujii et~al.}{2018}]{Fujii2018}
Fujii Y.,  et~al., 2018, \mn@doi [Astrobiology] {10.1089/ast.2017.1733}, 18,
  739

\bibitem[\protect\citeauthoryear{{Gaia Collaboration} et~al.,}{{Gaia
  Collaboration} et~al.}{2018}]{Gaia_Collaboration_2018}
{Gaia Collaboration} et~al., 2018, \mn@doi [\aap]
  {10.1051/0004-6361/201833051}, \href
  {https://ui.adsabs.harvard.edu/abs/2018A&A...616A...1G} {616, A1}

\bibitem[\protect\citeauthoryear{Gillon et~al.,}{Gillon
  et~al.}{2016}]{gillon_temperate_2016}
Gillon M.,  et~al., 2016, \mn@doi [Nature] {10.1038/nature17448}, 533, 221

\bibitem[\protect\citeauthoryear{{Gillon} et~al.,}{{Gillon}
  et~al.}{2017}]{Gillon2017}
{Gillon} M.,  et~al., 2017, \mn@doi [Nature] {10.1038/nature21360}, \href
  {https://ui.adsabs.harvard.edu/\#abs/2017Natur.542..456G} {542, 456}

\bibitem[\protect\citeauthoryear{{Gillon} et~al.,}{{Gillon}
  et~al.}{2020}]{Gillon_2020arXiv200204798G}
{Gillon} M.,  et~al., 2020, arXiv e-prints, \href
  {https://ui.adsabs.harvard.edu/abs/2020arXiv200204798G} {p. arXiv:2002.04798}

\bibitem[\protect\citeauthoryear{{Gordon} et~al.,}{{Gordon}
  et~al.}{2017}]{Gordon2017}
{Gordon} I.~E.,  et~al., 2017, \mn@doi [\jqsrt] {10.1016/j.jqsrt.2017.06.038},
  \href {https://ui.adsabs.harvard.edu/abs/2017JQSRT.203....3G} {203, 3}

\bibitem[\protect\citeauthoryear{{Grenfell}}{{Grenfell}}{2017}]{Grenfell_2017PhR...713....1G}
{Grenfell} J.~L.,  2017, \mn@doi [\physrep] {10.1016/j.physrep.2017.08.003},
  \href {https://ui.adsabs.harvard.edu/abs/2017PhR...713....1G} {713, 1}

\bibitem[\protect\citeauthoryear{{Grenfell}, {Gebauer}, {v. Paris}, {Godolt}
  \& {Rauer}}{{Grenfell} et~al.}{2014}]{Grenfell_2014P&SS...98...66G}
{Grenfell} J.~L.,  {Gebauer} S.,  {v. Paris} P.,  {Godolt} M.,   {Rauer} H.,
  2014, \mn@doi [\planss] {10.1016/j.pss.2013.10.006}, \href
  {https://ui.adsabs.harvard.edu/abs/2014P&SS...98...66G} {98, 66}

\bibitem[\protect\citeauthoryear{{Grimm} et~al.,}{{Grimm}
  et~al.}{2018}]{Grimm2018}
{Grimm} S.~L.,  et~al., 2018, \mn@doi [\aap] {10.1051/0004-6361/201732233},
  \href {https://ui.adsabs.harvard.edu/abs/2018A&A...613A..68G} {613, A68}

\bibitem[\protect\citeauthoryear{Günther et~al.,}{Günther
  et~al.}{2020}]{gunther_stellar_2020}
Günther M.~N.,  et~al., 2020, \mn@doi [AJ] {10.3847/1538-3881/ab5d3a}, 159, 60

\bibitem[\protect\citeauthoryear{{Hawker} \& {Parry}}{{Hawker} \&
  {Parry}}{2019}]{Hawker2019}
{Hawker} G.~A.,  {Parry} I.~R.,  2019, \mn@doi [\mnras] {10.1093/mnras/stz323},
  \href {https://ui.adsabs.harvard.edu/abs/2019MNRAS.484.4855H} {484, 4855}

\bibitem[\protect\citeauthoryear{{Johns}, {Marti}, {Huff}, {McCann},
  {Wittenmyer}, {Horner}  \& {Wright}}{{Johns}
  et~al.}{2018}]{Johns_2018ApJS..239...14J}
{Johns} D.,  {Marti} C.,  {Huff} M.,  {McCann} J.,  {Wittenmyer} R.~A.,
  {Horner} J.,   {Wright} D.~J.,  2018, \mn@doi [\apjs]
  {10.3847/1538-4365/aae5fb}, \href
  {https://ui.adsabs.harvard.edu/abs/2018ApJS..239...14J} {239, 14}

\bibitem[\protect\citeauthoryear{Kaltenegger}{Kaltenegger}{2017}]{kaltenegger_how_2017}
Kaltenegger L.,  2017, \mn@doi [Annu. Rev. Astron. Astrophys.]
  {10.1146/annurev-astro-082214-122238}, 55, 433

\bibitem[\protect\citeauthoryear{Kaltenegger \& Traub}{Kaltenegger \&
  Traub}{2009}]{kaltenegger_transits_2009}
Kaltenegger L.,  Traub W.~A.,  2009, \mn@doi [ApJ]
  {10.1088/0004-637X/698/1/519}, 698, 519

\bibitem[\protect\citeauthoryear{Kaltenegger, Traub  \& Jucks}{Kaltenegger
  et~al.}{2007}]{kaltenegger_spectral_2007}
Kaltenegger L.,  Traub W.~A.,   Jucks K.~W.,  2007, \mn@doi [ApJ]
  {10.1086/510996}, 658, 598

\bibitem[\protect\citeauthoryear{Kaltenegger, Lin  \& Madden}{Kaltenegger
  et~al.}{2020a}]{kaltenegger_high-resolution_2020}
Kaltenegger L.,  Lin Z.,   Madden J.,  2020a, \mn@doi [ApJ]
  {10.3847/2041-8213/ab789f}, 892, L17

\bibitem[\protect\citeauthoryear{{Kaltenegger}, {MacDonald}, {Kozakis},
  {Lewis}, {Mamajek}, {McDowell}  \& {Vanderburg}}{{Kaltenegger}
  et~al.}{2020b}]{Kaltenegger2020}
{Kaltenegger} L.,  {MacDonald} R.~J.,  {Kozakis} T.,  {Lewis} N.~K.,  {Mamajek}
  E.~E.,  {McDowell} J.~C.,   {Vanderburg} A.,  2020b, \mn@doi [\apjl]
  {10.3847/2041-8213/aba9d3}, \href
  {https://ui.adsabs.harvard.edu/abs/2020ApJ...901L...1K} {901, L1}

\bibitem[\protect\citeauthoryear{{Karman} et~al.,}{{Karman}
  et~al.}{2019}]{Karman2019}
{Karman} T.,  et~al., 2019, \mn@doi [\icarus] {10.1016/j.icarus.2019.02.034},
  \href {https://ui.adsabs.harvard.edu/abs/2019Icar..328..160K} {328, 160}

\bibitem[\protect\citeauthoryear{{Kasting} \& {Ackerman}}{{Kasting} \&
  {Ackerman}}{1986}]{Kasting_1986Sci...234.1383K}
{Kasting} J.~F.,  {Ackerman} T.~P.,  1986, \mn@doi [Science]
  {10.1126/science.234.4782.1383}, \href
  {https://ui.adsabs.harvard.edu/abs/1986Sci...234.1383K} {234, 1383}

\bibitem[\protect\citeauthoryear{{Kasting}, {Whitmire}  \&
  {Reynolds}}{{Kasting} et~al.}{1993}]{1993Icar..101..108K}
{Kasting} J.~F.,  {Whitmire} D.~P.,   {Reynolds} R.~T.,  1993, \mn@doi
  [\icarus] {10.1006/icar.1993.1010}, \href
  {https://ui.adsabs.harvard.edu/abs/1993Icar..101..108K} {101, 108}

\bibitem[\protect\citeauthoryear{{Kasting}, {Kopparapu}, {Ramirez}  \&
  {Harman}}{{Kasting} et~al.}{2014}]{Kasting_2014PNAS..11112641K}
{Kasting} J.~F.,  {Kopparapu} R.,  {Ramirez} R.~M.,   {Harman} C.~E.,  2014,
  \mn@doi [Proceedings of the National Academy of Science]
  {10.1073/pnas.1309107110}, \href
  {https://ui.adsabs.harvard.edu/abs/2014PNAS..11112641K} {111, 12641}

\bibitem[\protect\citeauthoryear{{Krissansen-Totton}, {Garland}, {Irwin}  \&
  {Catling}}{{Krissansen-Totton} et~al.}{2018}]{Krissansen-Totton2018}
{Krissansen-Totton} J.,  {Garland} R.,  {Irwin} P.,   {Catling} D.~C.,  2018,
  \mn@doi [\aj] {10.3847/1538-3881/aad564}, \href
  {https://ui.adsabs.harvard.edu/abs/2018AJ....156..114K} {156, 114}

\bibitem[\protect\citeauthoryear{Lammer et~al.,}{Lammer
  et~al.}{2007}]{lammer_coronal_2007}
Lammer H.,  et~al., 2007, \mn@doi [Astrobiology] {10.1089/ast.2006.0128}, 7,
  185

\bibitem[\protect\citeauthoryear{{Lederberg}}{{Lederberg}}{1965}]{Lederberg_1965Natur.207....9L}
{Lederberg} J.,  1965, \mn@doi [\nat] {10.1038/207009a0}, \href
  {https://ui.adsabs.harvard.edu/abs/1965Natur.207....9L} {207, 9}

\bibitem[\protect\citeauthoryear{{Lewis} et~al.,}{{Lewis}
  et~al.}{2020}]{Lewis2020}
{Lewis} N.~K.,  et~al., 2020, \mn@doi [\apjl] {10.3847/2041-8213/abb77f}, \href
  {https://ui.adsabs.harvard.edu/abs/2020ApJ...902L..19L} {902, L19}

\bibitem[\protect\citeauthoryear{Lin \& Kaltenegger}{Lin \&
  Kaltenegger}{2020}]{Lin2020}
Lin Z.,  Kaltenegger L.,  2020, \mn@doi [Monthly Notices of the Royal
  Astronomical Society] {10.1093/mnras/stz3213}

\bibitem[\protect\citeauthoryear{Lincowski, Meadows, Crisp, Robinson, Luger,
  Lustig-Yaeger  \& Arney}{Lincowski et~al.}{2018}]{lincowski_evolved_2018}
Lincowski A.~P.,  Meadows V.~S.,  Crisp D.,  Robinson T.~D.,  Luger R.,
  Lustig-Yaeger J.,   Arney G.~N.,  2018, \mn@doi [ApJ]
  {10.3847/1538-4357/aae36a}, 867, 76

\bibitem[\protect\citeauthoryear{Lingam \& Loeb}{Lingam \&
  Loeb}{2017}]{lingam_reduced_2017}
Lingam M.,  Loeb A.,  2017, \mn@doi [ApJ] {10.3847/2041-8213/aa8860}, 846, L21

\bibitem[\protect\citeauthoryear{{Lovelock}}{{Lovelock}}{1965}]{Lovelock_1965Natur.207..568L}
{Lovelock} J.~E.,  1965, \mn@doi [\nat] {10.1038/207568a0}, \href
  {https://ui.adsabs.harvard.edu/abs/1965Natur.207..568L} {207, 568}

\bibitem[\protect\citeauthoryear{Loyd et~al.,}{Loyd
  et~al.}{2018}]{loyd_muscles_2018}
Loyd R. O.~P.,  et~al., 2018, \mn@doi [ApJ] {10.3847/1538-4357/aae2bd}, 867, 71

\bibitem[\protect\citeauthoryear{{Lustig-Yaeger}, {Meadows}  \&
  {Lincowski}}{{Lustig-Yaeger} et~al.}{2019}]{Lustig-Yaeger2019}
{Lustig-Yaeger} J.,  {Meadows} V.~S.,   {Lincowski} A.~P.,  2019, \mn@doi [\aj]
  {10.3847/1538-3881/ab21e0}, \href
  {https://ui.adsabs.harvard.edu/abs/2019AJ....158...27L} {158, 27}

\bibitem[\protect\citeauthoryear{Lyons, Reinhard  \& Planavsky}{Lyons
  et~al.}{2014}]{lyons_rise_2014}
Lyons T.~W.,  Reinhard C.~T.,   Planavsky N.~J.,  2014, \mn@doi [Nature]
  {10.1038/nature13068}, 506, 307

\bibitem[\protect\citeauthoryear{{MacDonald} \& {Madhusudhan}}{{MacDonald} \&
  {Madhusudhan}}{2017}]{MacDonald2017}
{MacDonald} R.~J.,  {Madhusudhan} N.,  2017, \mn@doi [\mnras]
  {10.1093/mnras/stx804}, \href
  {https://ui.adsabs.harvard.edu/abs/2017MNRAS.469.1979M} {469, 1979}

\bibitem[\protect\citeauthoryear{{Madden} \& {Kaltenegger}}{{Madden} \&
  {Kaltenegger}}{2020}]{Madden_2020_surfaces}
{Madden} J.,  {Kaltenegger} L.,  2020, \mn@doi [\mnras]
  {10.1093/mnras/staa387}, \href
  {https://ui.adsabs.harvard.edu/abs/2020MNRAS.495....1M} {495, 1}

\bibitem[\protect\citeauthoryear{{Madhusudhan}}{{Madhusudhan}}{2018}]{madhusudhan2018}
{Madhusudhan} N.,  2018, {Atmospheric Retrieval of Exoplanets}.
p.~104, \mn@doi{10.1007/978-3-319-55333-7_104}

\bibitem[\protect\citeauthoryear{{Madhusudhan} \& {Seager}}{{Madhusudhan} \&
  {Seager}}{2009}]{Madhusudhan2009}
{Madhusudhan} N.,  {Seager} S.,  2009, \mn@doi [ApJ]
  {10.1088/0004-637X/707/1/24}, \href
  {https://ui.adsabs.harvard.edu/\#abs/2009ApJ...707...24M} {707, 24}

\bibitem[\protect\citeauthoryear{{Marley}, {Ackerman}, {Cuzzi}  \&
  {Kitzmann}}{{Marley} et~al.}{2013}]{Marley2013}
{Marley} M.~S.,  {Ackerman} A.~S.,  {Cuzzi} J.~N.,   {Kitzmann} D.,  2013,
  {Clouds and Hazes in Exoplanet Atmospheres}.
p.~367, \mn@doi{10.2458/azu\_uapress\_9780816530595-ch15}

\bibitem[\protect\citeauthoryear{{May}, {Taylor}, {Komacek}, {Line}  \&
  {Parmentier}}{{May} et~al.}{2021}]{May2021}
{May} E.~M.,  {Taylor} J.,  {Komacek} T.~D.,  {Line} M.~R.,   {Parmentier} V.,
  2021, arXiv e-prints, \href
  {https://ui.adsabs.harvard.edu/abs/2021arXiv210309313M} {p. arXiv:2103.09313}

\bibitem[\protect\citeauthoryear{{Moore} \& {Cowan}}{{Moore} \&
  {Cowan}}{2020}]{Moore2020}
{Moore} K.,  {Cowan} N.~B.,  2020, \mn@doi [\mnras] {10.1093/mnras/staa1796},
  \href {https://ui.adsabs.harvard.edu/abs/2020MNRAS.496.3786M} {496, 3786}

\bibitem[\protect\citeauthoryear{Moran, Hörst, Batalha, Lewis  \&
  Wakeford}{Moran et~al.}{2018}]{moran_limits_2018}
Moran S.~E.,  Hörst S.~M.,  Batalha N.~E.,  Lewis N.~K.,   Wakeford H.~R.,
  2018, \mn@doi [AJ] {10.3847/1538-3881/aae83a}, 156, 252

\bibitem[\protect\citeauthoryear{Morley, Kreidberg, Rustamkulov, Robinson  \&
  Fortney}{Morley et~al.}{2017}]{morley_observing_2017}
Morley C.~V.,  Kreidberg L.,  Rustamkulov Z.,  Robinson T.,   Fortney J.~J.,
  2017, \mn@doi [ApJ] {10.3847/1538-4357/aa927b}, 850, 121

\bibitem[\protect\citeauthoryear{{O'Malley-James} \&
  {Kaltenegger}}{{O'Malley-James} \&
  {Kaltenegger}}{2017}]{O'Malley-James_2017MNRAS.469L..26O}
{O'Malley-James} J.~T.,  {Kaltenegger} L.,  2017, \mn@doi [\mnras]
  {10.1093/mnrasl/slx047}, \href
  {https://ui.adsabs.harvard.edu/abs/2017MNRAS.469L..26O} {469, L26}

\bibitem[\protect\citeauthoryear{{Ohno} \& {Kawashima}}{{Ohno} \&
  {Kawashima}}{2020}]{Ohno2020}
{Ohno} K.,  {Kawashima} Y.,  2020, \mn@doi [\apjl] {10.3847/2041-8213/ab93d7},
  \href {https://ui.adsabs.harvard.edu/abs/2020ApJ...895L..47O} {895, L47}

\bibitem[\protect\citeauthoryear{O’Malley-James \&
  Kaltenegger}{O’Malley-James \&
  Kaltenegger}{2019}]{omalley-james_lessons_2019}
O’Malley-James J.~T.,  Kaltenegger L.,  2019, \mn@doi [Monthly Notices of the
  Royal Astronomical Society] {10.1093/mnras/stz724}, 485, 5598

\bibitem[\protect\citeauthoryear{{Robinson} et~al.,}{{Robinson}
  et~al.}{2011}]{Robinson_2011AsBio..11..393R}
{Robinson} T.~D.,  et~al., 2011, \mn@doi [Astrobiology]
  {10.1089/ast.2011.0642}, \href
  {https://ui.adsabs.harvard.edu/abs/2011AsBio..11..393R} {11, 393}

\bibitem[\protect\citeauthoryear{{Rodler} \& {L{\'o}pez-Morales}}{{Rodler} \&
  {L{\'o}pez-Morales}}{2014}]{Rodler2014}
{Rodler} F.,  {L{\'o}pez-Morales} M.,  2014, \mn@doi [\apj]
  {10.1088/0004-637X/781/1/54}, \href
  {https://ui.adsabs.harvard.edu/abs/2014ApJ...781...54R} {781, 54}

\bibitem[\protect\citeauthoryear{Rugheimer \& Kaltenegger}{Rugheimer \&
  Kaltenegger}{2018}]{rugheimer_spectra_2018}
Rugheimer S.,  Kaltenegger L.,  2018, \mn@doi [ApJ] {10.3847/1538-4357/aaa47a},
  854, 19

\bibitem[\protect\citeauthoryear{{Rugheimer}, {Kaltenegger}, {Zsom}, {Segura}
  \& {Sasselov}}{{Rugheimer} et~al.}{2013}]{Rugheimer_2013AsBio..13..251R}
{Rugheimer} S.,  {Kaltenegger} L.,  {Zsom} A.,  {Segura} A.,   {Sasselov} D.,
  2013, \mn@doi [Astrobiology] {10.1089/ast.2012.0888}, \href
  {https://ui.adsabs.harvard.edu/abs/2013AsBio..13..251R} {13, 251}

\bibitem[\protect\citeauthoryear{{Sagan}, {Thompson}, {Carlson}, {Gurnett}  \&
  {Hord}}{{Sagan} et~al.}{1993}]{Sagan_1993Natur.365..715S}
{Sagan} C.,  {Thompson} W.~R.,  {Carlson} R.,  {Gurnett} D.,   {Hord} C.,
  1993, \mn@doi [\nat] {10.1038/365715a0}, \href
  {https://ui.adsabs.harvard.edu/abs/1993Natur.365..715S} {365, 715}

\bibitem[\protect\citeauthoryear{Scalo et~al.,}{Scalo
  et~al.}{2007}]{scalo_m_2007}
Scalo J.,  et~al., 2007, \mn@doi [Astrobiology] {10.1089/ast.2006.0125}, 7, 85

\bibitem[\protect\citeauthoryear{{Schindler} \& {Kasting}}{{Schindler} \&
  {Kasting}}{2000}]{Schindler_2000Icar..145..262S}
{Schindler} T.~L.,  {Kasting} J.~F.,  2000, \mn@doi [\icarus]
  {10.1006/icar.2000.6340}, \href
  {https://ui.adsabs.harvard.edu/abs/2000Icar..145..262S} {145, 262}

\bibitem[\protect\citeauthoryear{{Schwieterman}, {Robinson}, {Meadows}, {Misra}
   \& {Domagal-Goldman}}{{Schwieterman} et~al.}{2015}]{Schwieterman2015}
{Schwieterman} E.~W.,  {Robinson} T.~D.,  {Meadows} V.~S.,  {Misra} A.,
  {Domagal-Goldman} S.,  2015, \mn@doi [\apj] {10.1088/0004-637X/810/1/57},
  \href {https://ui.adsabs.harvard.edu/abs/2015ApJ...810...57S} {810, 57}

\bibitem[\protect\citeauthoryear{{Schwieterman} et~al.,}{{Schwieterman}
  et~al.}{2018}]{Schwieterman_2018AsBio..18..663S}
{Schwieterman} E.~W.,  et~al., 2018, \mn@doi [Astrobiology]
  {10.1089/ast.2017.1729}, \href
  {https://ui.adsabs.harvard.edu/abs/2018AsBio..18..663S} {18, 663}

\bibitem[\protect\citeauthoryear{{Seager}, {Turner}, {Schafer}  \&
  {Ford}}{{Seager} et~al.}{2005}]{Seager_2005AsBio...5..372S}
{Seager} S.,  {Turner} E.~L.,  {Schafer} J.,   {Ford} E.~B.,  2005, \mn@doi
  [Astrobiology] {10.1089/ast.2005.5.372}, \href
  {https://ui.adsabs.harvard.edu/abs/2005AsBio...5..372S} {5, 372}

\bibitem[\protect\citeauthoryear{{Seager}, {Schrenk}  \& {Bains}}{{Seager}
  et~al.}{2012}]{Seager2012}
{Seager} S.,  {Schrenk} M.,   {Bains} W.,  2012, \mn@doi [Astrobiology]
  {10.1089/ast.2010.0489}, \href
  {https://ui.adsabs.harvard.edu/abs/2012AsBio..12...61S} {12, 61}

\bibitem[\protect\citeauthoryear{See, Jardine, Vidotto, Petit, Marsden, Jeffers
   \& do Nascimento}{See et~al.}{2014}]{see_effects_2014}
See V.,  Jardine M.,  Vidotto A.~A.,  Petit P.,  Marsden S.~C.,  Jeffers S.~V.,
    do Nascimento J.~D.,  2014, \mn@doi [A\&A] {10.1051/0004-6361/201424323},
  570, A99

\bibitem[\protect\citeauthoryear{{Segura}, {Kasting}, {Meadows}, {Cohen},
  {Scalo}, {Crisp}, {Butler}  \& {Tinetti}}{{Segura}
  et~al.}{2005}]{Segura_2005AsBio...5..706S}
{Segura} A.,  {Kasting} J.~F.,  {Meadows} V.,  {Cohen} M.,  {Scalo} J.,
  {Crisp} D.,  {Butler} R. A.~H.,   {Tinetti} G.,  2005, \mn@doi [Astrobiology]
  {10.1089/ast.2005.5.706}, \href
  {https://ui.adsabs.harvard.edu/abs/2005AsBio...5..706S} {5, 706}

\bibitem[\protect\citeauthoryear{{Serdyuchenko}, {Gorshelev}, {Weber},
  {Chehade}  \& {Burrows}}{{Serdyuchenko} et~al.}{2014}]{Serdyuchenko2014}
{Serdyuchenko} A.,  {Gorshelev} V.,  {Weber} M.,  {Chehade} W.,   {Burrows}
  J.~P.,  2014, \mn@doi [Atmospheric Measurement Techniques]
  {10.5194/amt-7-625-2014}, \href
  {https://ui.adsabs.harvard.edu/abs/2014AMT.....7..625S} {7, 625}

\bibitem[\protect\citeauthoryear{{Serindag} \& {Snellen}}{{Serindag} \&
  {Snellen}}{2019}]{Serindag2019}
{Serindag} D.~B.,  {Snellen} I. A.~G.,  2019, \mn@doi [\apjl]
  {10.3847/2041-8213/aafa1f}, \href
  {https://ui.adsabs.harvard.edu/abs/2019ApJ...871L...7S} {871, L7}

\bibitem[\protect\citeauthoryear{Shields, Ballard  \& Johnson}{Shields
  et~al.}{2016}]{shields_habitability_2016}
Shields A.~L.,  Ballard S.,   Johnson J.~A.,  2016, \mn@doi [Physics Reports]
  {10.1016/j.physrep.2016.10.003}, 663, 1

\bibitem[\protect\citeauthoryear{{Snellen}, {de Kok}, {le Poole}, {Brogi}  \&
  {Birkby}}{{Snellen} et~al.}{2013}]{Snellen2013}
{Snellen} I.~A.~G.,  {de Kok} R.~J.,  {le Poole} R.,  {Brogi} M.,   {Birkby}
  J.,  2013, \mn@doi [\apj] {10.1088/0004-637X/764/2/182}, \href
  {https://ui.adsabs.harvard.edu/abs/2013ApJ...764..182S} {764, 182}

\bibitem[\protect\citeauthoryear{{Toon}, {McKay}, {Ackerman}  \&
  {Santhanam}}{{Toon} et~al.}{1989}]{Toon_1989JGR....9416287T}
{Toon} O.~B.,  {McKay} C.~P.,  {Ackerman} T.~P.,   {Santhanam} K.,  1989,
  \mn@doi [\jgr] {10.1029/JD094iD13p16287}, \href
  {https://ui.adsabs.harvard.edu/abs/1989JGR....9416287T} {94, 16287}

\bibitem[\protect\citeauthoryear{{Traub} \& {Stier}}{{Traub} \&
  {Stier}}{1976}]{Traub_1976ApOpt..15..364T}
{Traub} W.~A.,  {Stier} M.~T.,  1976, \mn@doi [\ao] {10.1364/AO.15.000364},
  \href {https://ui.adsabs.harvard.edu/abs/1976ApOpt..15..364T} {15, 364}

\bibitem[\protect\citeauthoryear{{Tremblay}, {Line}, {Stevenson}, {Kataria},
  {Zellem}, {Fortney}  \& {Morley}}{{Tremblay} et~al.}{2020}]{Tremblay2020}
{Tremblay} L.,  {Line} M.~R.,  {Stevenson} K.,  {Kataria} T.,  {Zellem} R.~T.,
  {Fortney} J.~J.,   {Morley} C.,  2020, \mn@doi [\aj]
  {10.3847/1538-3881/ab64dd}, \href
  {https://ui.adsabs.harvard.edu/abs/2020AJ....159..117T} {159, 117}

\bibitem[\protect\citeauthoryear{{Trotta}}{{Trotta}}{2017}]{Trotta2017}
{Trotta} R.,  2017, arXiv e-prints, \href
  {https://ui.adsabs.harvard.edu/abs/2017arXiv170101467T} {p. arXiv:1701.01467}

\bibitem[\protect\citeauthoryear{{Turbet}, {Bolmont}, {Bourrier}, {Demory},
  {Leconte}, {Owen}  \& {Wolf}}{{Turbet} et~al.}{2020}]{Turbet2020}
{Turbet} M.,  {Bolmont} E.,  {Bourrier} V.,  {Demory} B.-O.,  {Leconte} J.,
  {Owen} J.,   {Wolf} E.~T.,  2020, \mn@doi [\ssr]
  {10.1007/s11214-020-00719-1}, \href
  {https://ui.adsabs.harvard.edu/abs/2020SSRv..216..100T} {216, 100}

\bibitem[\protect\citeauthoryear{{Wakeford} et~al.,}{{Wakeford}
  et~al.}{2017}]{Wakeford2017}
{Wakeford} H.~R.,  et~al., 2017, \mn@doi [Science] {10.1126/science.aah4668},
  \href {https://ui.adsabs.harvard.edu/abs/2017Sci...356..628W} {356, 628}

\bibitem[\protect\citeauthoryear{{Wilson} et~al.,}{{Wilson}
  et~al.}{2021}]{Wilson2021}
{Wilson} D.~J.,  et~al., 2021, \mn@doi [\apj] {10.3847/1538-4357/abe771}, \href
  {https://ui.adsabs.harvard.edu/abs/2021ApJ...911...18W} {911, 18}

\bibitem[\protect\citeauthoryear{Wunderlich et~al.,}{Wunderlich
  et~al.}{2019}]{wunderlich_detectability_2019}
Wunderlich F.,  et~al., 2019, \mn@doi [A\&A] {10.1051/0004-6361/201834504},
  624, A49

\bibitem[\protect\citeauthoryear{Wunderlich et~al.,}{Wunderlich
  et~al.}{2020}]{wunderlich_distinguishing_2020}
Wunderlich F.,  et~al., 2020, \mn@doi [The Astrophysical Journal]
  {10.3847/1538-4357/aba59c}, 901, 126

\bibitem[\protect\citeauthoryear{{Youngblood} et~al.,}{{Youngblood}
  et~al.}{2016}]{Youngblood_2016ApJ...824..101Y}
{Youngblood} A.,  et~al., 2016, \mn@doi [\apj] {10.3847/0004-637X/824/2/101},
  \href {https://ui.adsabs.harvard.edu/abs/2016ApJ...824..101Y} {824, 101}

\bibitem[\protect\citeauthoryear{Zahnle, Arndt, Cockell, Halliday, Nisbet,
  Selsis  \& Sleep}{Zahnle et~al.}{2007}]{zahnle_emergence_2007}
Zahnle K.,  Arndt N.,  Cockell C.,  Halliday A.,  Nisbet E.,  Selsis F.,
  Sleep N.~H.,  2007, \mn@doi [Space Sci Rev] {10.1007/s11214-007-9225-z}, 129,
  35

\bibitem[\protect\citeauthoryear{de Wit et~al.,}{de~Wit
  et~al.}{2018}]{de_wit_atmospheric_2018}
de Wit J.,  et~al., 2018, \mn@doi [Nat Astron] {10.1038/s41550-017-0374-z}, 2,
  214

\makeatother
\end{thebibliography}

\appendix

\section{Retrievals including Gaussian scatter}

\begin{figure*}
	\includegraphics[width=0.495\textwidth]{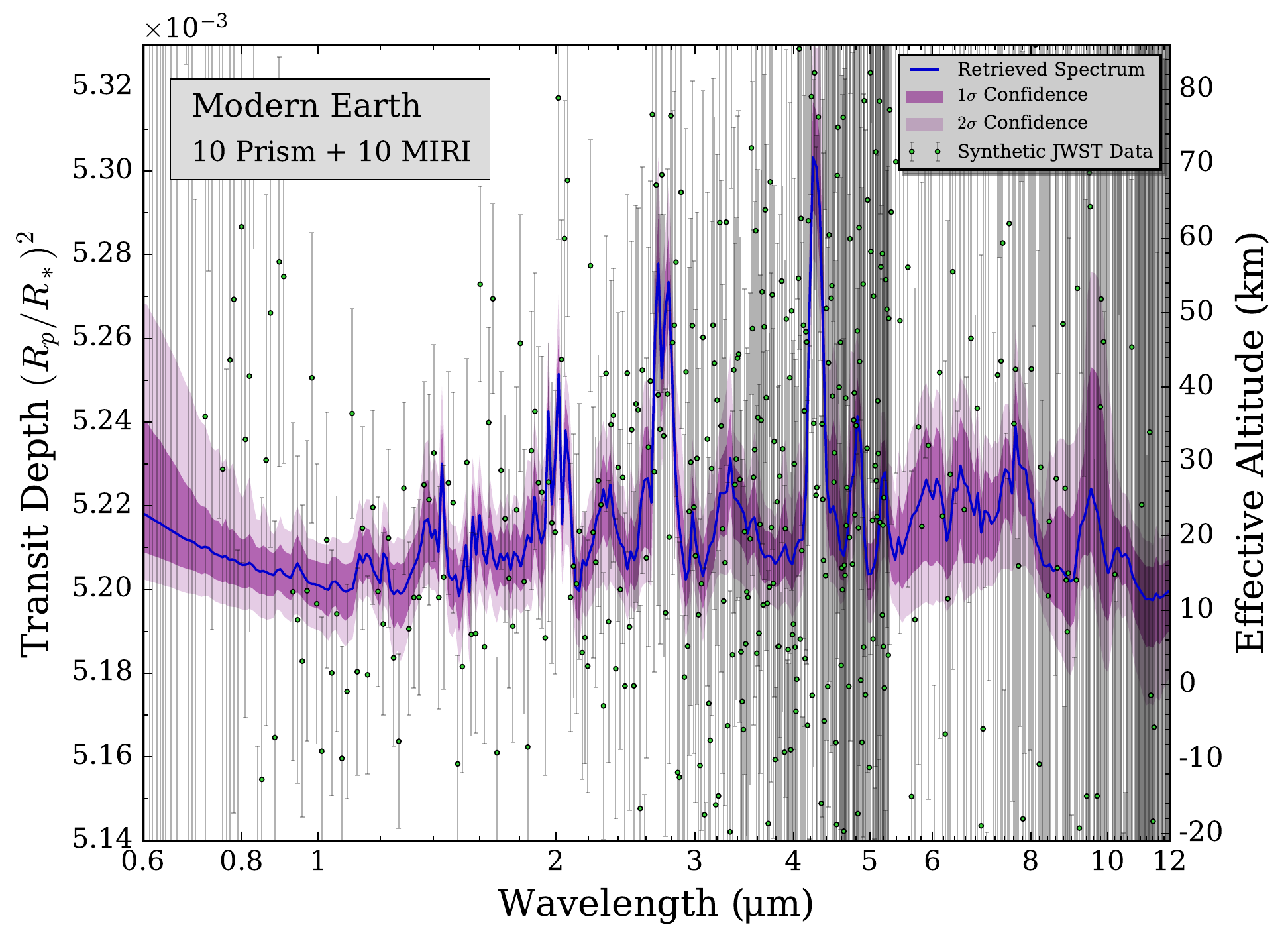}
	\includegraphics[width=0.495\textwidth]{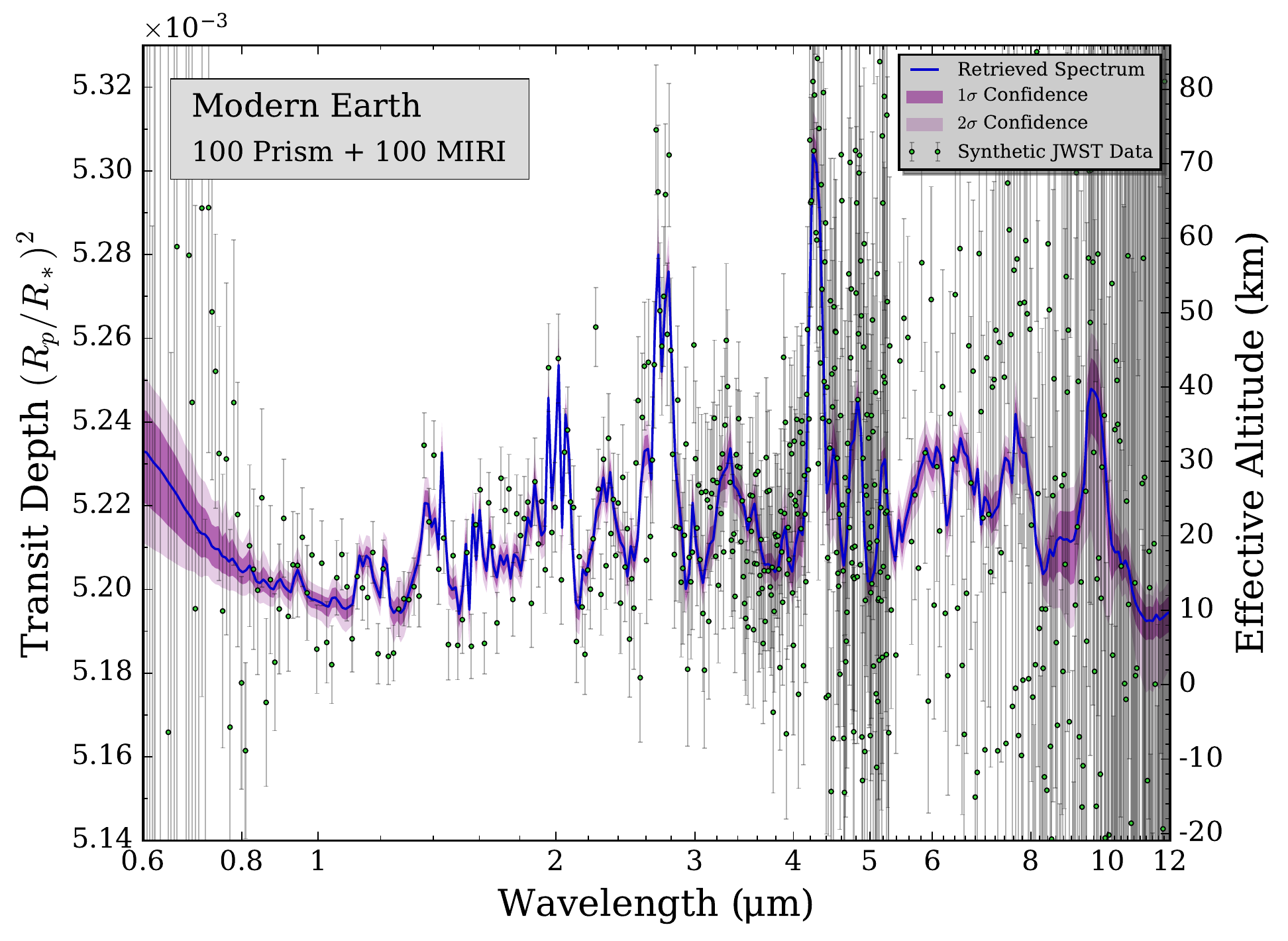}
	\includegraphics[width=0.495\textwidth]{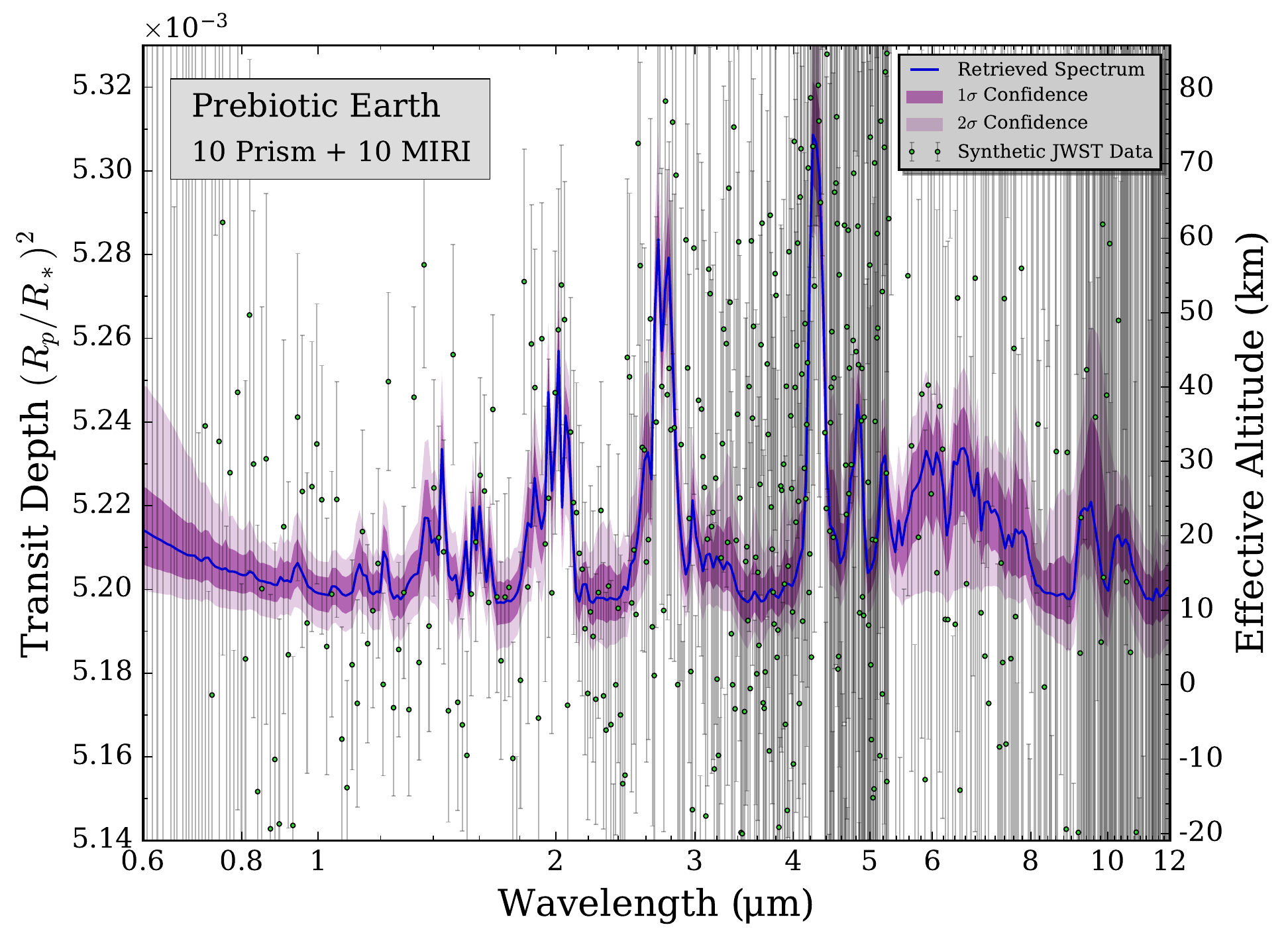}
	\includegraphics[width=0.495\textwidth]{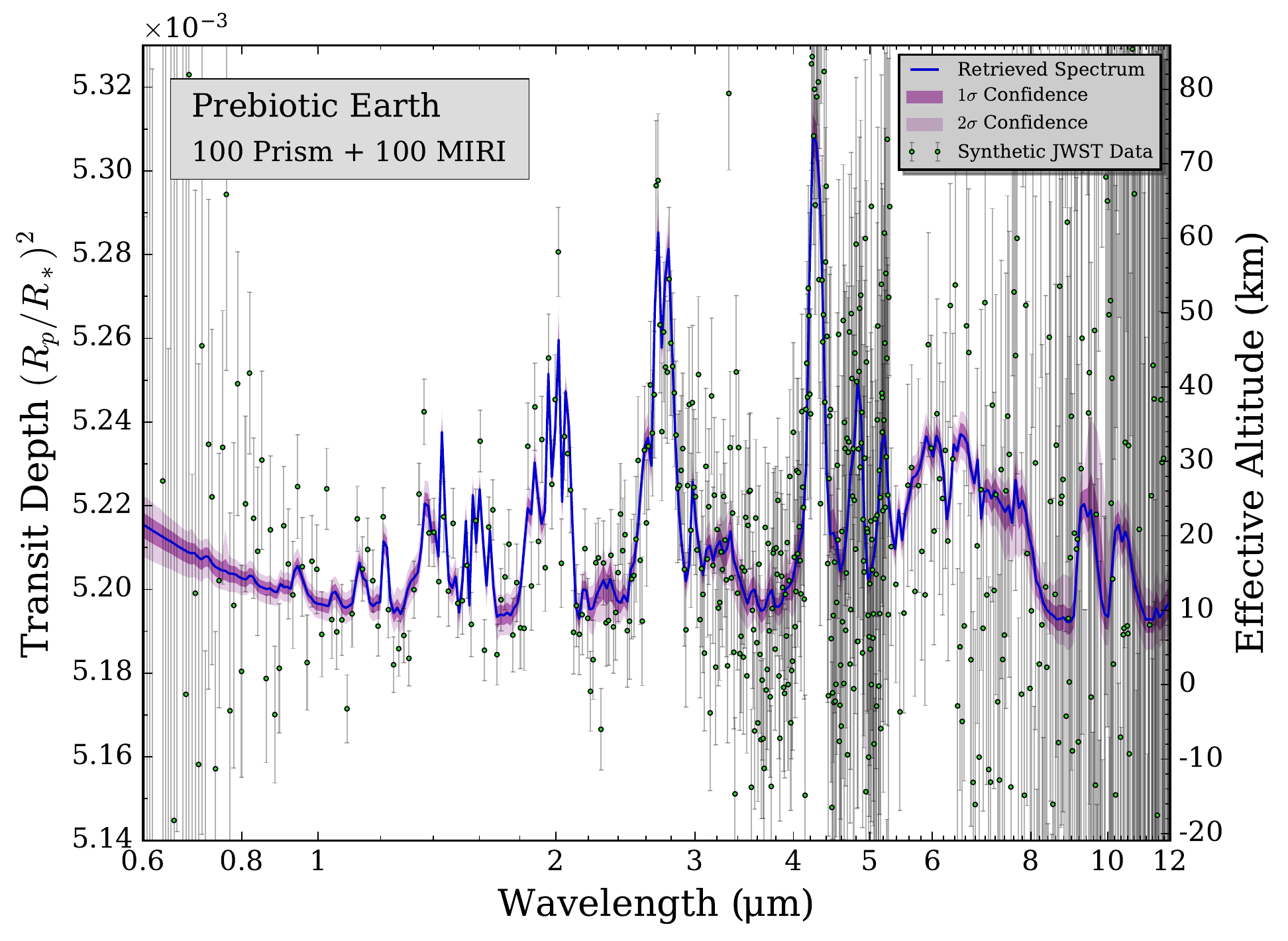}
    \caption{Retrieved transmission spectra for \textit{JWST} with Gaussian scatter. The four panels panels show the same cases depicted in Figure~\ref{fig:retrieved_spectrum}, namely 10 NIRSpec Prism + 10 MIRI LRS transits (left panels) and 100 NIRSpec Prism + 100 MIRI LRS transits (right panels) for our modern Earth (top panels) and prebiotic Earth (bottom panels) scenarios. The median retrieved spectra (blue) and 1$\sigma$ / 2$\sigma$ confidence intervals (purple contours) correctly capture the overall spectral morphology of the true model spectra (see Figure~\ref{fig:molecular_signatures}).}
    \label{fig:retrieved_spectra_Gauss_scatter}
\end{figure*}

Here we show retrieved transmission spectra from simulated \textit{JWST} datasets including Gaussian scatter. Figure~\ref{fig:retrieved_spectra_Gauss_scatter} shows simulated \textit{JWST} NIRSpec Prism and MIRI LRS data at the native instrument resolutions (i.e. without binning), highlighting the intrinsic scatter expected in transmission spectra of TRAPPIST-1e. Comparing our modern and prebiotic scenarios, we see hints of enhanced CH$_4$ features at 2.3$\,\micron$ and 3.4$\,\micron$ for the modern scenario, even with just 10 Prism transits. However, 100 transits with the Prism or MIRI are required before hints of O$_3$ are inferred for the modern scenario (via the Chappuis band or the 9.6$\,\micron$ feature).

\section{Retrieved CO$_2$ abundances}

\begin{figure*}
	\includegraphics[width=\textwidth]{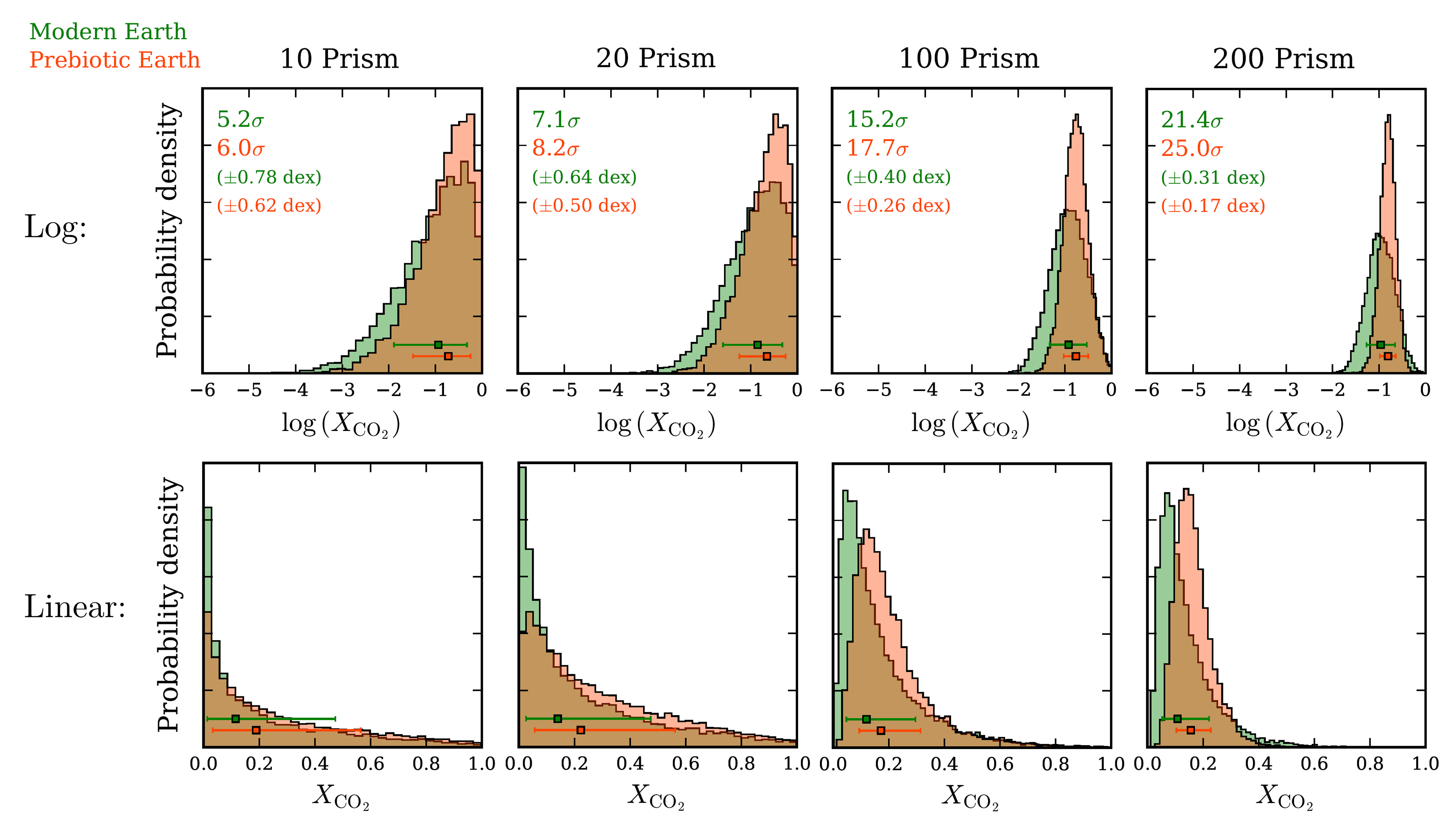}
    \caption{Retrieved CO$_2$ mixing ratios represented in logarithmic and linear space. The top row shows the retrieved CO$_2$ abundances for the modern Earth (green) and prebiotic Earth (orange) models as in Figure~\ref{fig:retrieved_abundances}. The bottom row shows the same retrieval results on a linear scale. The logarithmic scale demonstrates that lower limits can be placed on the CO$_2$ abundances in all cases, while the linear scale shows that upper limits can similarly be established. The latter result suggests that a CO$_2$ dominated atmosphere ($X_{\rm{CO_2}} = 1$) is disfavored even with just 10 NIRSpec Prism transits.}
    \label{fig:CO2_log_linear_comparison}
\end{figure*}

Here we provide an additional representation of our retrieved CO$_2$ abundances, focusing on the transition region between CO$_2$-rich ($X_{\rm{CO_2}} > 0.1$) to CO$_2$-dominated ($X_{\rm{CO_2}} \sim 1$) atmospheres. Figure~\ref{fig:CO2_log_linear_comparison} compares our retrieved CO$_2$ abundances on a logarithmic scale (as in Figure~\ref{fig:retrieved_abundances}) to the same abundances on a linear scale. Our retrievals correctly identify that CO$_2$ is an important constituent of TRAPPIST-1e's atmosphere, but that CO$_2$-dominated atmospheres are not the favored solution. Establishing the existence of a non-CO$_2$ background gas will be important to place the conditions on TRAPPIST-1e in context.

\bsp	
\label{lastpage}
\end{document}